\documentclass[
  pra,twocolumn,
  amsmath,
  amssymb,
  superscriptaddress,
]{revtex4}

\usepackage{bm}
\usepackage{graphicx}
\usepackage{color}

\def\be{\begin{equation}}
\def\ee{\end{equation}}
\def\bea{\begin{eqnarray}}
\def\eea{\end{eqnarray}}

\begin{document}

\title{Spectral diffusion and scaling of many-body delocalization transitions}

\author{I.V. Gornyi}
\affiliation{Institut f\"ur Nanotechnologie, Karlsruhe Institute of Technology, 76021 Karlsruhe, Germany}
 \affiliation{A.~F.\ Ioffe Physico-Technical Institute, 194021 St. Petersburg, Russia}
 \affiliation{\mbox{
 Institut f\"ur Theorie der Kondensierten Materie,
 Karlsruhe Institute of Technology, 76128 Karlsruhe, Germany}}
 \affiliation{L.~D.\ Landau Institute for Theoretical Physics RAS, 119334 Moscow, Russia}

\author{A.~D.\ Mirlin}
\affiliation{Institut f\"ur Nanotechnologie, Karlsruhe Institute of Technology, 76021 Karlsruhe, Germany}
\affiliation{\mbox{
 Institut f\"ur Theorie der Kondensierten Materie,
 Karlsruhe Institute of Technology, 76128 Karlsruhe, Germany}}
\affiliation{Petersburg Nuclear Physics Institute,  188300 St.~Petersburg, Russia}
\affiliation{L.~D.\ Landau Institute for Theoretical Physics RAS, 119334 Moscow, Russia}

\author{D.~G.\ Polyakov}
\affiliation{Institut f\"ur Nanotechnologie, Karlsruhe Institute of Technology, 76021 Karlsruhe, Germany}

\author{A.L. Burin}
\affiliation{Department of Chemistry, Tulane University, New Orleans, LA 70118, USA}

\begin{abstract}
We analyze the role of spectral diffusion in the problem of many-body delocalization in quantum dots
and in extended systems. The spectral diffusion parametrically enhances delocalization,
modifying the scaling of the delocalization threshold with the interaction coupling constant.
\end{abstract}


\maketitle

\section{Introduction}
\label{s1}

The essence of the phenomenon of Anderson localization \cite{anderson58} is conventionally understood in terms of spatial localization of a quantum particle by disorder in the absence of interactions of the particle with other particles, or any other degrees of freedom in the system for that matter. A key feature of Anderson localization is the sharp boundary between continua of localized and delocalized single-particle states,
as the particle energy is varied. For a macroscopic noninteracting electron system, this means a quantum phase transition at temperature $T=0$, namely the Anderson transition between the metallic and insulating phases, with changing, e.g., Fermi energy. Localization theory has been under intensive development up until recently, culminating in a rich variety of the universality classes of the Anderson transition depending on the underlying symmetries and topologies~\cite{evers08}. For the generic case of a random scalar potential, however, all single-particle states, independently of their energy, are localized for $d=1$, where $d$ is the space dimension, and (in the absence of a magnetic field leading to the quantum Hall effect and of spin-orbit coupling) for $d=2$, the latter being the lower critical dimensionality for the metal-insulator transition for this type of disorder.

One of the key questions in this area is how the inelastic transitions due to electron-electron interactions affect Anderson localization at nonzero $T$. More specifically, one may wonder about the resulting $T$ dependence of the electrical conductivity $\sigma (T)$ in the phase that is insulating at zero $T$. It was about a third of a century ago that two important advances were made in order to answer the latter question. On the one hand, it was argued in Ref.~\cite{fleishman80} that short-range interactions in a disordered system with localized single-particle states do not destroy localization also at nonzero $T$, implying that $\sigma(T)=0$ in the absence of a zero-$T$ mobility edge, at least in the strongly disordered limit. The reasoning was based on a perturbation series analysis of the tentative single-particle decay through the excitation of localized particle-hole pairs.

On the other hand, it was found in Ref.~\cite{altshuler82} that inelastic electron-electron scattering in the weakly disordered limit leads to a finite dephasing rate and thus to a nonzero $\sigma(T)$, even for the case of $d\leq 2$ where all single-particle states are localized. The calculation rested on a self-consistent description of Nyquist noise produced by delocalized electrons. The latter approach offered an appealing picture of the weak-localization limit for disordered Fermi liquids \cite{altshuler85}.

Assuming that the two arguments~\cite{fleishman80,altshuler82}, taken at face value, are correct, a disordered interacting system with localized single-particle states must undergo a finite-$T$ localization-delocalization transition with changing strength of disorder. If disorder is not too strong (the condition relevant to tight-binding lattice models), the localization-delocalization transition must also occur as $T$ is changed. In other words, at certain finite $T$, inelastic scattering due to electron-electron interactions
should lead to a delocalization (``many-body delocalization'') of electrons that are localized in the absence of interactions. At that time, however, and up until the last decade, this logical construction, although obviously having fundamental theoretical implications, was not commonly appreciated as an essential part of the conceptual framework for disorder-induced phase transitions. Important early ideas in this direction (in particular, the very possibility of a finite-$T$ localization-delocalization transition) were put forward in Ref.~\cite{kagan84}.

Interest in the problem was revived in the context of Anderson localization in Fock space. The concept of viewing higher-order electron-electron scattering as the Anderson localization problem on a certain graph (cf. Ref.~\cite{abouchacra73}) in Fock space was proposed in Ref.~\cite{altshuler97} with the aim to study the hybridization of a highly excited single-particle state, as a function of its energy, in a quantum dot. This concept has become a central element in subsequent developments, see Ref.~\cite{gornyi16} for the recent overview.

While Ref.~\cite{altshuler97} dealt with a closed zero-dimensional system, subsequent works addressed
the problem of localization in a spatially extended many-body system at finite $T$, with one of the goals being
to resolve the dichotomy between the early results of Refs.~\cite{fleishman80} and \cite{altshuler82}.
Soon after, two papers~\cite{gornyi05,basko06}, studying the interaction-induced decay of a single-particle state for the short-ranged interaction in the spirit of Refs.~\cite{fleishman80} and \cite{altshuler97}, obtained  the scaling of the ratio of a higher-order matrix element to the relevant many-body level spacing with the order of the perturbation theory (equivalently, with the ``generation number'' in Fock space). Both papers~\cite{gornyi05,basko06} arrived at the conclusion that there is a finite-$T$ localization-delocalization transition, with $\sigma(T)$ being zero on the low-$T$ side of the transition and nonzero on the other side.
The result was obtained by means of a mapping on the Bethe-lattice problem  in the vicinity of the transition in Ref.~\cite{gornyi05} and a self-consistent treatment of the perturbation series in Ref.~\cite{basko06}.

The articles \cite{gornyi05,basko06} initiated a considerable interest in the problem of many-body localization at nonzero temperature in spatially extended systems. Numerous works studied the problem by means of numerical simulations of one-dimensional systems in the presence of disorder and (short-range) interaction, see, in particular, Refs.~\cite{oganesyan07,monthus10,kjall14,gopalakrishnan14,luitz15,nandkishore15,karrasch15,imbrie16}, with results supporting the existence of the transition.  An analytical proof of the existence of many-body localization for a particular spin-chain model was proposed in Ref.~\cite{imbrie16a}.
On the experimental side, signatures of a many-body localization-delocalization transition
in interacting systems were observed  in indium-oxide films~\cite{ovadyahu,ovadia15} as well as
in a $d=1$ system of fermionic atoms placed in two incommensurate optical lattices~\cite{schreiber15} and in a $d=2$ system of cold bosonic atoms~\cite{choi16}. Many-body localization was studied experimentally also in arrays of coupled one-dimensional
optical lattices~\cite{bordia15,lueschen16}.

The main goal of this paper is to revisit the scaling of the position of the many-body delocalization transition (in terms of a critical strength of disorder or a critical temperature) with the strength of (weak) interaction in a broad class of models specified in Sec. \ref{s2} below.
These include a quantum dot with interacting electrons, a spin quantum dot, as well as spatially extended systems of spins or electrons (with localized single-particle states) with short-range interactions.

Identifying the position of the many-body delocalization transition in many-body systems is a highly nontrivial problem. Indeed, a naive Thouless criterion that compares a matrix element with the level spacing of directly coupled states may be very deceptive. For example, in an extended system with localized single-particle states the level spacing of many-body states coupled to a given typical many-body state is inversely proportional to the system volume $L^d$, where $L$ is the linear size of the system. Employing the above naive criterion, one would arrive at the conclusion that in the thermodynamic limit $L\to \infty$ all many-body states are extended for an arbitrarily weak (but finite) interaction and at arbitrarily low temperature $T\ne 0$, in contradiction with Refs.~\cite{fleishman80,gornyi05,basko06}. The problem with the above argumentation becomes particularly obvious if one notices that it could also be applied to the situation in which different localization volumes are fully decoupled, and thus many-body states are definitely localized.

In Refs.~\cite{gornyi05,basko06}, as a solution to this problem, it was proposed to
consider single-particle excitations. Specifically, following essentially the arguments of Refs.~\cite{fleishman80} and \cite{altshuler97}, the criterion for the many-body localization-delocalization transition was identified (up to a logarithmic factor)
by comparing the typical interaction matrix element with the
typical level spacing of final (three-particle: two electrons and a hole) states
for the decay of a single-electron excitation. The obtained position of the
transition was determined by the critical condition for a
Bethe lattice~\cite{abouchacra73} in Fock space~\cite{altshuler97}.

While the approaches used in Refs.~\cite{gornyi05,basko06} were different,
the results for the transition temperature $T_c$ are the same, up to a numerical
prefactor, see Ref.~\cite{ros15} for a recent detailed discussion. Specifically,
it was found~\cite{gornyi05,basko06} that the system exhibits a transition between
the low-temperature localized phase~\cite{fleishman80} and the high-temperature
delocalized phase~\cite{altshuler82} at the temperature
\begin{equation}
\label{e1}
T_c \sim \frac{\Delta_\xi}{\alpha\ln(1/\alpha)}~,
\end{equation}
where $\Delta_\xi$ is the characteristic single-particle level spacing in the localization volume, $\alpha\ll 1$ is the dimensionless strength of the short-range interaction, and the symbol ``$\sim$'' means ``up to a numerical coefficient of order unity''.

Recently, in Ref.~\cite{gornyi16}, an analysis of the localization-delocalization transition in an electronic quantum dot \cite{altshuler97} was performed that is a direct counterpart of the analysis of a spatially extended system  in Refs.~\cite{gornyi05} and \cite{basko06}. Specifically, it was found \cite{gornyi16} that many-body states in a quantum dot undergo the Fock-space delocalization transition at the energy
\begin{equation}
\label{e2}
E_c=(g/\ln g)\Delta,
\end{equation}
which can be translated into an effective transition temperature~\cite{gornyi16}
\begin{equation}
\label{e3}
T_c\sim (E_c \Delta)^{1/2} \sim \Delta(g/\ln g)^{1/2}~.
\end{equation}
Here, $\Delta$ is the characteristic single-electron level spacing in the dot, and $g\gg 1$ is the dimensionless conductance which determines the characteristic value of the interaction matrix elements
$V\sim \Delta/g$ [thus $1/g$ plays a role of $\alpha$ in Eq.~(\ref{e1})].
The threshold (\ref{e3}) emerges in a particularly transparent way when one considers the decay of a single-particle excitation on top of a thermal state in a quantum dot.
In full analogy with Eq.~(\ref{e1}), this result can be obtained (up to a logarithmic factor) by comparing the matrix element with the level spacing of directly coupled states.

In the present paper, we argue that the approximation adopted in the previous works \cite{gornyi05,basko06,ros15,gornyi16} discarded an important ingredient that affects the scaling of the delocalization threshold: \textit{spectral diffusion}. This relaxation mechanism was identified in the theory of spectral lines in spin-resonance phenomena~\cite{klauder62} and was later analyzed in the context of the relaxation properties of glasses in Refs.~\cite{black77,galperin83,burin90,burin95}.
In particular, in Refs.~\cite{burin90} and \cite{burin95}, the importance of spectral diffusion for delocalization
of collective excitations was emphasized.
The essence of spectral diffusion is that transitions that have already taken place in an
interacting many-body system shift resonant conditions for other possible transitions.
Recently, it was shown that spectral diffusion plays an essential role in determining the position of the many-body delocalization transition in a finite system with power-law interactions \cite{burin15a,gutman16} (see also Ref.~\cite{kucsko16} where manifestations of spectral diffusion  in this context were observed experimentally).
The aim of the present work is to show that a similar mechanism leads to enhancement of many-body delocalization in quantum dots and in extended systems with short-range interaction, and to determine the correct scaling of the delocalization thresholds with the parameters of the system.

On the technical level, Refs.~\cite{gornyi05,basko06,ros15,gornyi16} neglected diagonal (in the single-particle basis) terms of the interaction,
retaining only the matrix elements of interaction with all four indices being distinct.
The usual argument behind this approximation refers to the Hartree-Fock basis that
includes such diagonal terms on the single-particle level. We will show below, however, that such an approximation is, in fact, too naive:
the spectral diffusion driven by diagonal matrix elements changes the localization-delocalization threshold parametrically.

When addressing extended systems in this work, we consider a model with finite-range interaction.
At the same time, it is worth noting that, while the relevant matrix elements of the long-range Coulomb interaction are expected to be effectively short-ranged in one-dimensional systems, the Coulomb interaction in higher-dimensional (2D and 3D) systems was shown to lead to the delocalization of excitations at
arbitrary temperatures \cite{burin89,burin95,Burin98,burin06,yao14,gutman16}. More specifically,
the delocalization occurs when the spatial dimensionality exceeds the half of the exponent
of the power-law decay of random long-range interactions.

The remainder of the paper is organized as follows. In Sec.~\ref{s2}, we describe the models of interest and
remind the reader of previous results for these models.
In Sec.~\ref{s3}, we show that the Fock-space delocalization in a quantum dot is substantially enhanced by spectral diffusion and determine the scaling of the corresponding delocalization threshold. In Sec.~\ref{s4}, we analyze the role of spectral diffusion in the problem of many-body delocalization in extended systems.
In Sec.~\ref{s6}, we summarize our findings, discuss further implications of our work, and outline possible generalizations of our theory.

\section{Models of the many-body delocalization transition}
\label{s2}

In this Section we list four (closely interrelated) models. The scaling of many-body delocalization transitions in these models---with the emphasis on the role of spectral diffusion---is the main subject of this work.

\subsection{Electron quantum dot}
\label{s2A}

\subsubsection{Model}
\label{s2A1}

We begin by considering a disordered quantum dot with the mean single-particle level spacing $\Delta$ (i.e., with the single-particle density of states $\nu_1=1/\Delta$) and dimensionless conductance $g\gg 1$. The Thouless energy is given by
\begin{equation}
E_{T}\sim g \Delta.
\label{ETh}
\end{equation}
For simplicity, we focus on the system of spinless (spin-polarized) electrons.
The Hamiltonian of the dot reads
\begin{equation}
H=\sum \epsilon_i c^\dagger_i c_i + \frac{1}{2}\sum V_{ijkl} c^\dagger_i c^\dagger_j c_k c_l,
\label{Hamiltonian}
\end{equation}
where $\epsilon_i$ are the energies of single-particle orbitals.
These orbitals are coupled by a two-particle interaction with matrix elements $V_{ijkl}$ being random quantities.  The root-mean-square (r.m.s.) deviation of the matrix elements for the Coulomb interaction (screened by electrons in the quantum dot) behaves as~\cite{blanter96,aleiner01}
\be
V\sim\Delta/g
\label{V}
\ee
when all energies of the involved single-particle states belong to the energy band of width $E_T$ around the Fermi level. When the difference between the energies of single-particle states
exceeds the Thouless energy, the matrix element is suppressed in a power-law manner as a function of the energy difference.
In what follows,
we will neglect such matrix elements. In fact, for the quantum-dot problem, states with energies above $E_T$ will be of no interest anyway, since the localization threshold is below $E_T$. The decay of the matrix elements will, however, become important for the case of spatially extended systems, see Sec.~\ref{s2B} below.

Let us consider a typical state with the total energy $E$.
The characteristic number of excited quasiparticles forming this state is
\be
N \sim (E/\Delta)^{1/2}.
\label{N}
\ee
and the characteristic energy of each particle (``effective temperature'') is
\be
T\sim (E\Delta)^{1/2} \sim N \Delta~.
\label{T}
\ee

For a given dimensionless interaction strength $1/g \ll 1$, the system undergoes a Fock-space delocalization transition with increasing energy. The question of the scaling of the transition is how the critical value $N_c \gg 1$ (or, equivalently, the critical temperature $T_c \sim N_c\Delta$, or the critical energy $E_c \sim N_c^2 \Delta$) scales with $1/g$.

\subsubsection{Previous result: ``Conservative estimate'' for the delocalization transition}
\label{s2A2}

We briefly remind the reader about the result for $N_c$ that is obtained \cite{gornyi16} if one neglects the effect of the shift of single-particle levels. More specifically, one considers the decay of a single-particle excitation, keeping contributions corresponding to creation of a maximum number of quasiparticles at each order of the perturbation theory. This is a direct counterpart of the ``forward approximation'' used in the context of spatially extended systems in Refs.~\cite{gornyi05,basko06,ros15}, see  Sec.~\ref{s2B2}. The level spacing of three-particle states to which a single-particle state with energy $\sim T$
is connected scales as
\be
\Delta_3(T)\sim\Delta^3/T^2 \sim \Delta^2/E.
\label{delta3}
\ee
Equating this to $V$, Eq.~(\ref{V}), one gets the energy
\be
E \sim g\Delta.
\label{E1}
\ee
Taking into account higher-order processes allows one to gain a logarithmic factor (familiar from the Bethe-lattice problem), yielding
\be
E_c^{\rm nsd} \sim (g/\ln g) \Delta,
\label{E1log}
\ee
or, equivalently,
\be
T_c^{\rm nsd} \sim N_c^{\rm nsd}\Delta \sim  (g/\ln g)^{1/2} \Delta.
\label{T1log}
\ee
This is the result for the delocalization border as found in Ref.~\cite{gornyi16}. The superscript ``nsd'' indicates that this result was obtained within the approximation that neglects spectral diffusion.
In view of what comes below, one can consider it as a ``conservative approximation'':
the states above this energy are definitely delocalized. In what follows, we will improve this result by including the effects of spectral diffusion which will parametrically lower the critical energy.

\subsection{Electrons with spatially localized single-particle states and weak short-range interaction}
\label{s2B}

\subsubsection{Model}
\label{s2B1}

Let us now specify the model used in the study of many-body delocalization in
an extended system of fermions with localized single-particle states.
Consider the basis of localized (with the localization length $\xi$)
single-particle states $\phi_\alpha$ and a
short-range interaction between them. The Hamiltonian of the system reads:
\begin{equation}
H=\sum_\alpha \epsilon_\alpha c^\dagger_\alpha c_\alpha
+\frac{1}{2} \sum V_{\alpha\beta\gamma\delta}
c^\dagger_\alpha c^\dagger_\beta c_\gamma c_\delta.
\label{ham}
\end{equation}
The interaction term here is antisymmetrized: $V_{\alpha\beta\gamma\delta}=V_{\beta\alpha\delta\gamma}
=-V_{\beta\alpha\gamma\delta}=-V_{\alpha\beta\delta\gamma}$.

 The matrix elements $V_{\alpha\beta\gamma\delta}$ of the short-range interaction connect only those states that are within a distance $\sim \xi $ in real space, where $\xi$ is the single-particle localization length. (We will assume that $\xi$ is of the same order for all single-particle states of interest.) Furthermore, only those matrix elements are retained that couple electron-hole pairs with an energy difference $\lesssim \Delta_\xi$, where  $\Delta_\xi\sim 1/\rho\xi^d$ is the single-particle level spacing in the localization volume, $\rho$ is the single-particle density of states:
 \be
\label{energy-condition}
|\epsilon_\alpha-\epsilon_\gamma|, \
|\epsilon_\beta-\epsilon_\delta|\lesssim \Delta_\xi \quad {\rm or} \quad
|\epsilon_\alpha-\epsilon_\delta|, \
|\epsilon_\beta-\epsilon_\gamma|\lesssim \Delta_\xi.
\ee
For larger energy differences, the matrix elements are suppressed and can be neglected
(cf. the matrix elements in a quantum dot with the dimensionless conductance set to $g\sim 1$).
The retained matrix elements are of the order of
\be
\label{e5}
V_{\alpha\beta\gamma\delta} \sim V \sim \alpha\Delta_\xi,
\ee
where $\alpha$ is the interaction constant. We will assume the case of weak interaction,  $\alpha \ll 1$.

With increasing temperature, the system undergoes a many-body delocalization transition at a certain critical temperature $T_c$. The question of the scaling of the transition threshold is how  the dimensionless ratio $T_c/\Delta_\xi \gg 1$ scales with the dimensionless coupling constant $\alpha \ll 1$.

\subsubsection{Previous result: ``Conservative estimate'' for the delocalization transition}
\label{s2B2}

Similarly to Sec.~\ref{s2A2}, we remind the reader about the result for the scaling of the transition as obtained in previous works, Refs.~\cite{gornyi05,basko06,ros15}. The approximation used there (termed ``forward approximation'' in Ref.~\cite{ros15}) was in keeping only diagrams with all quasiparticles involved being in different states. The arguments in favor of this approximation look at first sight quite convincing. First, one can argue that matrix elements with repeated indices are taken into account by choosing the Hartree-Fock single-particle states, see Sec. 2.2 of Ref.~\cite{basko06}. The second argument is essentially that the number of such matrix elements is parametrically smaller, so that they do not affect the combinatorial estimates of the probabilities to encounter a resonance at higher order, see Refs.~\cite{gornyi05,ros15}. While we will show in this paper that these arguments are in fact misleading, we present here the results of Refs.~\cite{gornyi05,basko06,ros15} obtained within the above approximation.

Consider the perturbative expansion for the decay of a localized single-particle excitation  $|\alpha\rangle$
 due to the interaction term in the Hamiltonian (\ref{ham}).
The lowest-order process is the transition to a
three-particle state -- two electrons $|\beta\rangle$,
$|\gamma\rangle$ and a hole $|\delta\rangle$, all located
within a distance $\sim \xi$.
Since energies of all the relevant single-particle states are within the
window of width $\sim T$ and satisfy the restriction (\ref{energy-condition}), the level
spacing of those three-particle states to which the original state
$|\alpha\rangle$  is coupled according to (\ref{e5}), reads
\be
\label{e6}
\Delta_\xi^{(3)} \sim \Delta_\xi^2/T.
\ee
Using Eqs.~(\ref{e5}) and (\ref{e6}), one finds the golden-rule
inelastic broadening of single-particle states
\be
\label{e7}
\Gamma\sim  |V|^2/\Delta_\xi^{(3)}
\sim \alpha^2 T.
\ee
The natural condition of validity of the golden-rule calculation is $V \gg \Delta_\xi^{(3)}$, or,
equivalently,  $\Gamma \gg \Delta_\xi^{(3)}$, which translates into $T\gg T_3$, where
\begin{equation}
T_3=\Delta_\xi/\alpha.
\label{T3}
\end{equation}
Below this temperature the elementary decay processes described by the lowest-order matrix element (\ref{e5}) are typically blocked by the energy conservation. The analysis of higher-order processes which takes
into account resonant transitions between distant states in Fock space \cite{gornyi05,basko06,ros15} yields the
critical temperature which differs from Eq.~(\ref{T3}) only logarithmically,
\begin{equation}
\label{Tc-ext-nsd}
T_c^{\rm nsd} \sim \frac{\Delta_\xi}{\alpha\ln(1/\alpha)}.
\end{equation}
where the superscript ``nsd'' again  indicates the neglect of the spectral diffusion.

\subsection{Spin quantum dot}
\label{s2C}

The analysis of the electronic model of Sec.~\ref{s2B} suggests introducing a closely related spin model. Specifically, let us combine states within one localization volume in pairs according to their energy ordering, and then replace each pair by a spin 1/2. We then get the following model
\be
\label{H-spin-quantum-dot}
H = \sum_{k,k'=1}^N \sum_{i,j}v_{kk'}^{ij} S_k^i S_{k'}^j + \sum_{k=1}^N h_k S_k^z,
\ee
where $S_k$ are spin-1/2 operators, $i,j$ are cartesian components (taking values $x,y,z$), $v_{kk'}^{ij}$ are random interactions with zero
 mean and the r.m.s. value
 \be
 \label{v-variance}
V=\left \langle \left( v_{kk'}^{ij}\right)^2 \right \rangle^{1/2}  = \alpha \Delta,
 \ee
 and $h_k$ is a random Zeeman field with a uniform (box) distribution in the range $0 < h < \Delta$.

 At infinite temperature, $T = \infty$, the problem is characterized by two dimensionless parameters: the number of spins $N \gg 1$ and the interaction strength $\alpha \ll 1$. At fixed $\alpha$ and with increasing $N$, the system undergoes a many-body delocalization transition at a certain critical value $N_c$. The question of the scaling of the transition threshold is how  $N_c$ scales with $\alpha \ll 1$.

A correspondence between the parameters of this ``spin quantum dot'' model and a single localization volume of the electronic model of Sec.~\ref{s2B1} is quite transparent. Specifically, the scale $\Delta$ of the present model corresponds to $\Delta_\xi$ of Sec.~\ref{s2B1}, and the number of states $N$ of the present model (considered at infinite $T$) corresponds to $T/\Delta_\xi$ of Sec.~\ref{s2B1}. The interaction strength is denoted by $\alpha$ in both cases, to emphasize the correspondence. While the two models are clearly not identical, they are expected to share the same scaling of the transition, as discussed in detail below.

 \subsection{Extended spin system with weak finite-range interaction}
\label{s2D}

Having introduced in Sec.~\ref{s2C} the spin quantum dot model corresponding to a
single localization volume of the electronic model of Sec.~\ref{s2B}, we promote it to a spatially extended system (cf., e.g., Ref.~\cite{burin90}). This generalization is quite straightforward: we arrange such spin quantum dots in a $d$-dimensional lattice and introduce interaction between spins in adjacent quantum dots of the same order as the interaction between spins within the same quantum dot. The Hamiltonian thus reads
\be
\label{H-spin-extended}
H = \sum_{\langle k,k' \rangle}\sum_{i,j}v_{kk'}^{ij} S_k^i S_{k'}^j + \sum_kh_k S_k^z,
\ee
where the summation in the first term goes over pairs of spins belonging to the same box or to the adjacent boxes. The distributions of the corresponding interaction matrix elements $v_{kk'}^{ij}$ and of  the random fields $h_k$ are the same as in the single-quantum-dot problem, Sec.~\ref{s2C}.

In full analogy with the case of a single spin quantum dot, the problem is characterized, at infinite temperature, $T = \infty$, by two dimensionless parameters: the number of spins in each quantum dot, $N \gg 1$, and the interaction strength $\alpha \ll 1$. Again, at fixed $\alpha$ and with increasing $N$, the system undergoes a many-body delocalization transition at a certain critical value $N_c$.
The question of the scaling of the transition threshold is the same: how  $N_c$ scales
with $\alpha \ll 1$.

A correspondence between the parameters of this spatially extended spin model and the electronic model of Sec.~\ref{s2B1} is established in the same way as for the problem of spin quantum dot. Again, the scale $\Delta$ of the present model corresponds to $\Delta_\xi$ of Sec.~\ref{s2B1}, the number of states $N$ in each quantum dot in the present model (considered at infinite $T$) corresponds to $T/\Delta_\xi$ of Sec.~\ref{s2B1}, and the interaction strength is $\alpha$ in both cases. The models are clearly not identical: in particular, in the electronic model one can speak about transport of charge and of energy, while in the spin model only energy transport can be considered. Nevertheless, the scaling of the many-body delocalization threshold, $N_c(\alpha)$, is expected to be the same in two models, as will be discussed below.

\section{Enhancement of delocalization by spectral diffusion in quantum dots}
\label{s3}

We begin the analysis of the scaling of delocalization transitions by considering the quantum dot problems defined in Secs.~\ref{s2A} and \ref{s2C}. Later we will generalize the obtained results onto the spatially extended models of Secs.~\ref{s2B} and \ref{s2D}.

\subsection{Electron quantum dot}
\label{s3A}

\subsubsection{First-order resonances: Delocalization down to $N \sim g^{2/5}$}
\label{s3A1}

We start with the model of an electronic quantum dot defined in Sec.~\ref{s2A1}.
Consider a given many-body basis state represented by a (Hartree-Fock) Slater determinant
with $N$ quasiparticles, see Eq.~\ref{N}. According to the conservative estimate presented in Sec.~\ref{s2A2}, the position of the threshold was determined (up to a logarithmic factor) by the requirement that each of $N$ involved quasiparticles has of the order of unity resonance decay channels in the first order of perturbation theory (i.e., basis states resonantly coupled to the original one by a first-order process involving the given particle). In other words, this requirement is equivalent
to the condition that a typical basis state is resonantly connected to $\sim N$ other
basis states. Each such resonance corresponds to a process $\alpha,\beta \to \gamma,\delta$,
where the states $\alpha$ and $\beta$ are occupied in the given basis state,
while the states $\gamma$ and $\delta$ are empty, and the energies satisfy
\be
\label{resonance-condition}
|\epsilon_\alpha + \epsilon_\beta - \epsilon_\gamma - \epsilon_\delta| \lesssim V.
\ee
This has led to the estimate (\ref{E1log}) for the threshold. (At the moment we will discard logarithmic factors and will return to them in the end of the analysis.) We will now show that the above condition for delocalization is in fact parametrically too strong: it is sufficient that there are in total $\sim 1$ resonances for a typical basis state. The key role in this enhancement of delocalization is played by spectral diffusion.

The level spacing of all basis   states to which a given  basis state is connected is
\begin{equation}
\Delta^{(N)}\sim   \frac{\Delta}{N^3} \sim \Delta \left(\frac{\Delta}{E}\right)^{3/2}~.
\label{DeltaNtyp}
\end{equation}
[We remind the reader that the connection between the energy $E$ and the number of particles $N$ is provided by Eq.~(\ref{N}).]
The condition $\Delta^{(N)} \sim V$ yields the energy at which resonances start to appear, i.e., a typical many-body basis state starts to hybridize with its Fock-space neighbors:
\begin{equation}
N \sim g^{1/3}, \quad {\rm or, equivalently,} \quad  E \sim g^{2/3}\Delta.
\label{g13}
\end{equation}
This energy scale was identified in Refs.~\cite{jacquod97,mirlin97,silvestrov98}.
It is also worth mentioning that the
relevance of the energy scale at which $\Delta^{(N)}\sim V$ to the onset of quantum chaos in nuclei was conjectured (on the basis of numerical simulations) by {\AA}berg \cite{aberg} who considered
a random-interaction model similar to the quantum-dot model defined in Sec.~\ref{s2A1}.

\begin{figure*}
  \centering
  \includegraphics[width=\textwidth]{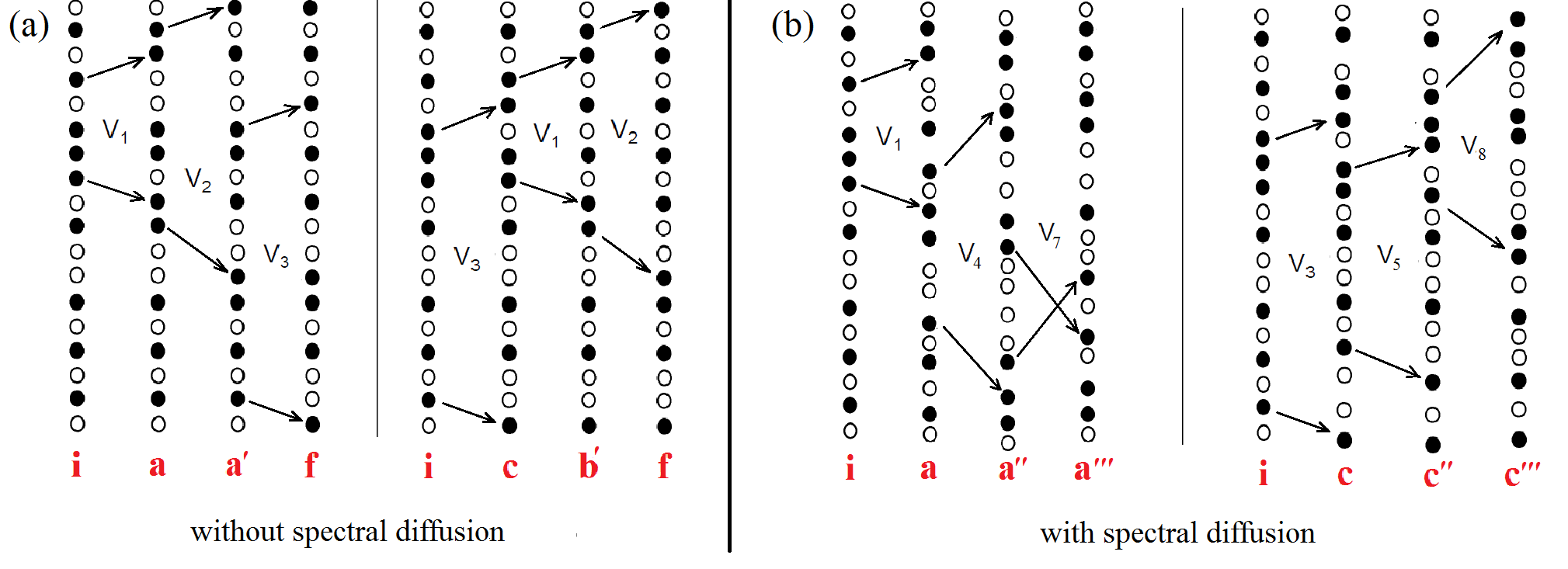}
  \caption{Schematic illustration of resonant processes for $N=6$, $p=3$ starting from the initial many-body state $i$ in the absence (a) and presence (b) of spectral diffusion. Full (empty) dots denote filled (empty) single-particle states. Without spectral diffusion (a), there are only three elementary resonances mediated by $V_1$, $V_2$, and $V_3$; all $p!=6$ possible consecutive orderings of these resonances produce the same final state $f$, such that the corresponding amplitudes interfere. In contrast, spectral diffusion (b), emphasized by random shifts of the original levels, leads to a possibility of reinitialization of resonances at each step, so that new resonances mediated by $V_4$, $V_5$, $V_7$, and $V_8$ appear, leading to distinct final states $a^{\prime\prime\prime}$ and $c^{\prime\prime\prime}$. The corresponding resonant networks
  in Fock space are shown in Fig.~\ref{Fig:Fock}.}
  \label{Fig:dots}
\end{figure*}

\begin{figure}
  \centerline{
  \includegraphics[width=8.5cm]{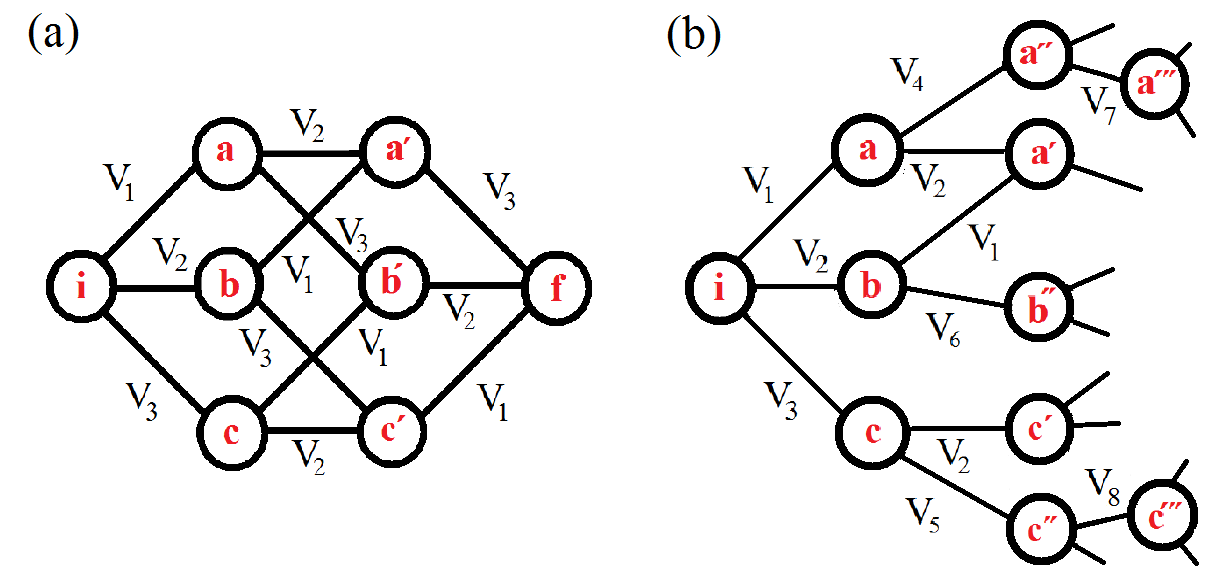}}
  \caption{Resonant networks in Fock space without (a) and with (b) spectral diffusion. The paths $i\to a\to a^\prime\to f$
  and $i\to c\to c^\prime\to f$ in panel (a) are schematically depicted in Fig.~\ref{Fig:dots}(a). The paths $i\to a\to a^{\prime\prime}\to a^{\prime\prime\prime}$
  and $i\to c \to c^{\prime\prime}\to c^{\prime\prime\prime}$ in panel (b) are schematically depicted in Fig.~\ref{Fig:dots}(b).  }
  \label{Fig:Fock}
\end{figure}

Consider a basis state with the energy $E$ above the threshold (\ref{g13}):
\begin{equation}
g^{2/3}\Delta \lesssim E \lesssim  g\Delta.
\end{equation}
Equivalently, this means that the energy per particle $T$ and the number of particles $N \sim T/\Delta$ satisfy
\begin{equation}
g^{1/3}\Delta \lesssim T  \lesssim g^{1/2}\Delta, \qquad g^{1/3} \lesssim N  \lesssim g^{1/2}.
\end{equation}
The number of resonances for such a many-body basis state is estimated as
\begin{equation}
p \sim \frac{V}{\Delta^{(N)}}\sim \frac{N^3}{g}.
\label{p}
\end{equation}
In terms of an effective tight-binding model in the Fock space, this means that the considered state (represented by a site on a graph in such a representation) is well (resonantly) connected to $p$ further states, Fig.~\ref{Fig:dots}(a). Each of these $p$ states is in turn well connected to $p$ states, etc. One could thus conclude that the problems maps essentially on a tree-like graph (looking locally as a Bethe lattice) with $p$ well-coupled neighbors for each vertex, which would imply a delocalization transition at $p \sim 1$. (Recall once more that we neglect for a while logarithms originating from higher-order resonant processes.) The problem with this argument is that the resonances that we encounter at higher orders are, as was pointed out in Ref.~\cite{silvestrov97},
not necessarily new ones, see a discussion in
Refs.~\cite{silvestrov98,basko06,ros15,gornyi16}. In particular, if the shift of energy levels due to transitions of other electrons (which is the source of spectral diffusion) is discarded \cite{burin15b}, the hybridization will only be able to proceed up to generation $p$, at which stage all resonances will be exhausted, see Fig.~\ref{Fig:Fock}(a). As a result, our original basis state will hybridize with only $\sim 2^p$  further basis states.
A full hybridization of states in the many-body Hilbert space requires that one can proceed
up to generation $N$. Equating $p=N$, one obtains (up to a logarithmic factor) the ``conservative estimate'' (\ref{E1log}), (\ref{T1log}).

Now we are ready to incorporate the spectral diffusion. For this purpose, we should take into account the effect of diagonal matrix elements $V_{\alpha\beta\beta\alpha} = - V_{\alpha\beta\alpha\beta}$ whose fluctuations~\cite{aleiner01}
have the same magnitude $V$, Eq.~(\ref{V}), as those of nondiagonal elements (the average diagonal matrix elements give a constant correction to the total energy and hence are not important in the present context). At each step two electrons change their states.  As a result of this, and because of
the diagonal matrix elements, energies of all other electrons are shifted by a random amount $\sim V$.  This means that roughly a half of $p$ resonances will not satisfy any more the resonant condition at the next step. Since the number $p$ of resonances is essentially the same for all typical basis states, these resonances will be replaced by new ones,
see Fig.~\ref{Fig:dots}(b). Thus, the spectral diffusion leads to a kind of ``reinitialization'' of resonances.

How far can we proceed now with constructing the resonance network  in Fock space (Fig.~\ref{Fig:Fock})?  Let us assume that we proceed up to generation $m$. During this process, the spectral diffusion shifts all energies by a random amount of the order of $m^{1/2}V$. In other words, the sets of single-particle levels $(\alpha,\beta; \gamma,\delta)$ that will provide a resonant step in such a process will be typically those that have originally a mismatch $|\epsilon_\alpha + \epsilon_\beta - \epsilon_\gamma - \epsilon_\delta| \sim m^{1/2}V$. The total number $n_m$ of sets $(\alpha,\beta; \gamma,\delta)$ in the energy window $m^{1/2}V$ is
$$
n_m\sim \frac{m^{1/2}V}{\Delta^{(N)}}.
$$
As long as this number is large compared with $m$, we will have at each step $\simeq p$ resonances that have not been ``used'' at previous steps, i.e., the process will proceed up to generation $m$ without a problem. In this regime, the effective Fock-space structure illustrated in Fig.~\ref{Fig:Fock}(b) has a lot of similarity with a tight-binding model on a tree-like graph; we will return to this analogy below. The maximum generation $m$ up to which we can proceed is thus given by the condition
\begin{equation}
m \Delta^{(N)} \sim m^{1/2}V,
\label{spec-diff-1}
\end{equation}
which yields
\begin{equation}
\label{m}
m \sim p^2.
\end{equation}
This number restricts the size of the cluster in Fig.~\ref{Fig:Fock}(b).

We thus have shown that the spectral diffusion allows one to proceed
to a parametrically larger resonance generation ($m \sim p^2$ instead of $m\sim p$). Equating $m =  N$ and using Eq.~(\ref{p}) for $p$, we get the number of electrons $N$ (and thus the energy $E$)
above which all electrons will be involved in resonances within the considered mechanism:
\begin{equation}
N \sim g^{2/5}, \qquad E \sim g^{4/5}\Delta.
\label{E45}
\end{equation}
Therefore, the actual delocalization border is in fact parametrically lower than the estimates (\ref{E1log}), (\ref{T1log}) discarding the spectral diffusion.

Let us analyze the scaling of the relaxation rate in the range
\be
\label{spectral-diff1}
g^{2/5} \lesssim N \lesssim g^{1/2}, \ \  {\rm or, equivalently,}  \ \  g^{4/5}\Delta \lesssim E \lesssim g \Delta
\ee
covered by the above spectral-diffusion mechanism. The total decay rate of a many-body basis state
reads
\begin{equation}
\Gamma_\text{MB}\sim \frac{V^2}{\Delta^{(N)}}\sim \frac{\Delta^2/g^2}{\Delta_3/N}\sim \Delta\frac{N^3}{g^2}.
\end{equation}
Thus, the relaxation rate for a given single-particle excitation can be estimated as
\begin{equation}
\label{GR1}
\frac{1}{\tau} \sim \frac{1}{N}\Gamma_\text{MB} \sim \Delta\frac{N^2}{g^2},
\end{equation}
which is nothing but the conventional golden-rule formula
\begin{equation}
\label{GR2}
\frac{1}{\tau} \sim \frac{V^2}{\Delta_3} \sim \frac{E}{g^2}.
\end{equation}

\begin{figure}
  \centerline{
  \includegraphics[width=9cm]{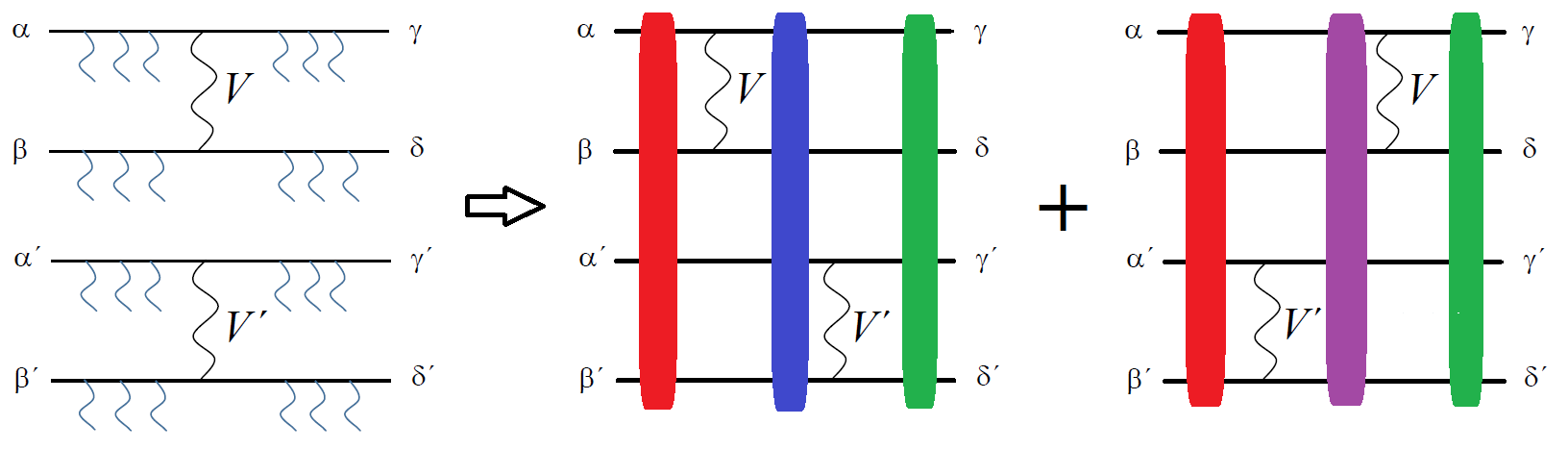}}
  \caption{Second-order process. Short wavy lines denote the diagonal interactions between the participating particles.
  The composite matrix element is a sum of two contributions that differ by the ordering of the off-diagonal interactions $V$ and $V^\prime$. The colored blocks denote the coupling of the particles by the diagonal interactions. The middle blocks representing the virtual states in the two contributions differ from each other
  and hence include different diagonal matrix elements.
  The change of energy between the virtual state (blue block) and initial state in the first term is given by
  $\omega=\epsilon+U_{\gamma\alpha^\prime}+U_{\gamma\beta^\prime}+U_{\delta\alpha^\prime}+U_{\delta\beta^\prime}
  -U_{\alpha\alpha^\prime}-U_{\alpha\beta^\prime}-U_{\beta\alpha^\prime}-U_{\beta\beta^\prime}$, whereas in the second term (with the magenta block) it is  $\omega^\prime=\epsilon^\prime+U_{\gamma^\prime\alpha}+U_{\gamma^\prime\beta}+U_{\delta^\prime\alpha}+U_{\delta^\prime\beta}
  -U_{\alpha\alpha^\prime}-U_{\alpha\beta^\prime}-U_{\beta\alpha^\prime}-U_{\beta\beta^\prime}$.
  Here we use the short-hand notation $U_{\mu\nu}=V_{\mu\nu\mu\nu}$ for the diagonal interactions.
  }
  \label{Fig:2res}
\end{figure}

This is not the end of the story, though. Up to now, we have only considered the
lowest-order resonant processes.
Including higher-order processes (see Fig.~\ref{Fig:2res}), we will show below that the delocalization border is in fact still lower---in the sense of its power-law dependence on the effective interaction strength $1/g$---than Eq.~(\ref{E45}). However, before proceeding with the analysis of higher-order resonance processes, it is worthwhile to pause for a moment and to analyze why the previous arguments suggesting that the diagonal matrix elements can be safely discarded have turned out to be incorrect.

\subsubsection{Discussion: Why diagonal matrix elements do matter}
\label{s3A2}

In Sec.~\ref{s2B2} we described the earlier arguments~\cite{gornyi05,basko06,ros15} suggesting that the diagonal matrix elements do not affect the scaling of the transition. Since we have shown that the inclusion of these matrix elements parametrically enhances the delocalization via spectral diffusion,
it is natural to discuss where the loopholes in the previous arguments were.

The first argument was that the diagonal matrix elements can be absorbed into the definition of Hartree-Fock single-particle levels. It is clear, however, that the Hartree-Fock energies depend on the occupation of other states. It is exactly the variation of the Hartree-Fock energies due to repopulation of other states that leads to spectral diffusion.

The second argument was essentially of a counting type: there are much less of diagonal matrix elements than nondiagonal ones. So, it seems that, when contributions to the decay rate of single-particle excitations are estimated in the perturbation theory, the diagonal terms can be neglected. Here two points should be emphasized. First, including the shift of Hartree-Fock energies amounts to inserting ladders between every pair of involved quasiparticles, which is already a resummation of an infinite series from the point of view of Feynman diagrammatics, see Sec.~\ref{s4C} below. Second, apart from the insertion of ladders, the relaxation of a single-particle excitation requires,  in the regime where the spectral diffusion is important for delocalization, considering processes of parametrically higher order that describe redistribution of other quasiparticles. Due to a combination of these two reasons, it is difficult to straightforwardly include spectral diffusion in the analysis of a decay of single-particle excitations. This explains why its effect was missed in Refs.~\cite{gornyi05,basko06,ros15}. In particular, this invalidates the proof of stability of the insulating phase at temperatures $T < T_c^{\rm nsd}$ presented in Ref.~\cite{basko06}: as we see now, the insulating phase might become stable only at parametrically lower temperatures.

\subsubsection{Higher-order processes: Delocalization down to $N \sim g^{1/3}$}
\label{s3A3}

Consider now a second-order transition resulting from a combination of two first-order processes, $\alpha,\beta \to \gamma,\delta$ and $\alpha',\beta' \to \gamma',\delta'$, see Fig.~\ref{Fig:2res}. We assume that each of the constituent first-order processes is non-resonant (virtual), i.e., the associated energy mismatches $\epsilon, \epsilon'$ are  large, $|\epsilon|, |\epsilon'| \gg V$, whereas the total second-order process is resonant. The energy mismatch associated with this process is
\be
\label{epsilon2}
\epsilon^{(2)} = \epsilon + \epsilon' + U,
\ee
where $U \sim V$ originates from the diagonal interactions between the particles of the set $\alpha,\beta,\gamma,\delta$ and those of the set $\alpha',\beta',\gamma',\delta'$ (Fig.~\ref{Fig:2res}). The origin of $U$ is thus the same as that of the spectral diffusion.

The effective matrix element of the transition is found by using the second order of the perturbation theory,
\be
\label{V2}
V^{(2)} = V_{\alpha\beta\gamma\delta} V_{\alpha'\beta'\gamma'\delta'} \left( \frac{1}{\epsilon} + \frac{1}{\epsilon'}\right) \sim
V^2 \frac {\epsilon + \epsilon'}{\epsilon\epsilon'}.
\ee
Since the resonance condition requires $|\epsilon^{(2)}| \ll V$ (see below), we find from Eq.~(\ref{epsilon2}) that  $|\epsilon + \epsilon'| \simeq |U| \sim V$. Therefore, the amplitude of the considered second-order resonant transition is estimated as
\be
\label{V2-estimate}
|V^{(2)}| \sim \frac{V^3}{|\epsilon\epsilon'|} \sim  \frac{V^3}{\epsilon^2}.
\ee
It is worth emphasizing an important role that was played by the term $U$ in  Eq.~(\ref{epsilon2}) for this estimate. Without this term, Eq.~(\ref{V2-estimate}) would give a vanishing result at resonance, $\epsilon^{(2)}=0$, which is a well-known cancelation that has been discussed, in particular, in Refs.~\cite{silvestrov97,silvestrov98,basko06,ros15,gornyi16}. These are diagonal matrix elements that couple two processes, $\alpha,\beta \to \gamma,\delta$ and $\alpha',\beta' \to \gamma',\delta'$, thus yielding a contribution $U$ to Eq.~(\ref{epsilon2}) and destroying the cancelation.

We analyze now whether such second-order resonant processes are sufficient to ensure delocalization in the whole many-body system.
In this analysis, we will consider the intermediate-state energy $\epsilon$ as a free parameter and will optimize with respect to it (a more detailed calculation is presented in Appendix~\ref{App:1}).
The delocalization requires fulfilment of two conditions. First, there should be resonances at each step, which means that the effective matrix element should be larger than the level spacing of final states $\Delta^{(N)}_{(2)}$ (to be evaluated below),
\be
\label{res-2}
|V^{(2)}| \gtrsim \Delta^{(N)}_{(2)}.
\ee
Second, the spectral diffusion should be sufficiently efficient to ensure that the process can proceed up to a generation $\sim N$. The corresponding condition is a direct generalization of Eq.~(\ref{spec-diff-1}):
\be
\label{spec-diff-2}
N \Delta^{(N)}_{(2)} \lesssim N^{1/2} V.
\ee
The level spacing $\Delta^{(N)}_{(2)}$ is found to be
\be
\label{Delta-N-2}
\Delta^{(N)}_{(2)} \sim \frac{\left(\Delta^{(N)}\right)^2}{\epsilon} \sim \frac{\Delta^2}{N^6\epsilon},
\ee
if the energy of the intermediate state is restricted to the interval between $\epsilon$ and $2\epsilon$.
Indeed, if we do not impose any restrictions on the intermediate state, we have typically $\epsilon \sim N \Delta$ and the level spacing of final states $\Delta^{(N)}_{(2)} \sim \Delta/N^7$. Restricting the energy of the intermediate state to be of order $\epsilon$, we reduce the number of possible finite states proportionally to $\epsilon$, which leads to an increase of   $\Delta^{(N)}_{(2)}$ proportionally to $1/\epsilon$.

Using Eq.~(\ref{Delta-N-2}), we rewrite the conditions (\ref{res-2}) and (\ref{spec-diff-2}) in the form
\be
\label{res-2a}
\epsilon \lesssim \frac{\Delta N^6}{g^3}
\ee
and
\be
\label{spec-diff2a}
\epsilon \gtrsim \frac{\Delta g}{N^{11/2}},
\ee
respectively. These conditions are compatible for $N \gtrsim N_c^{(2)}$, where
\be
\label{Nc2}
N_c^{(2)} \sim g^{8/23}.
\ee
Thus, taking into account the second-order processes (in combination with spectral diffusion)
has allowed us to further lower the boundary of the region for which the delocalization has been shown from Eq.~(\ref{E45}) to Eq.~(\ref{Nc2}).

This argumentation can be extended to processes of still higher orders.
Consider resonant transitions of $k$-th order that go through virtual states with energies $\sim \epsilon$.
The composite matrix element describing the transition of $2k$ particles mediated by $k$ elementary
off-diagonal matrix elements and fully dressed by diagonal matrix elements can be written as (cf. Ref.~\cite{burin16}):
\begin{eqnarray}
V^{(k)}=V_k\sum_{\{t\}_k}\prod_{i=1}^{k-1} \frac{V_i}{\sum_{j=1}^{i}\tilde{\epsilon}_{t_j}+\sum_{j'<j\leq i}\tilde{U}_{t_j t_{j'}}}.
\label{Vk-exact}
\end{eqnarray}
Here the summation is performed over all $k!$ permutations $\{t\}_k$ of the orderings of the off-diagonal matrix elements
$V_l=V_{\alpha_l\beta_l\gamma_l\delta_l}$ and
$\tilde{\epsilon}_{l}$ denotes the energy mismatch for the $l$-th pair (with inclusion of the diagonal matrix elements
of interaction with the ``background''). The term $\tilde{U}_{lm}$ in the denominator describes the diagonal interactions between
$l$-th and $m$-th pairs: it is a linear combination of the diagonal matrix elements $V_{{\mu_l}{\mu_m}{\mu_l}{\mu_m}}$
between the states belonging to the pairs $l$ and $m$, see Fig.~\ref{Fig:2res}.

To count resonances, it is convenient to introduce the ``coupling constant''
$\eta_k$ (see Appendix \ref{App:1}) which is a ratio of the composite matrix element $V^{(k)}$ and
the energy mismatch $\epsilon^{(k)}$ between the initial and final many-body states.
The latter is given by the same sum
of all $\tilde{\epsilon}$ and $\tilde{U}$ for each of the orderings $\{t\}_k$ of the off-diagonal matrix elements $V_j$:
\begin{equation}
\epsilon^{(k)}=\sum_{j=1}^k \tilde{\epsilon}_j + \sum_{j'<j\leq k}\tilde{U}_{j j'}.
\end{equation}
In the limit $U\gg V$ (large diagonal energy shifts \cite{burin16}), the coupling constant reduces to the one on a Bethe lattice.
For $U=0$, the sum over $k!$ permutations transforms identically to a single term \cite{silvestrov97,silvestrov98,basko06,ros15,gornyi16}.
For example, the explicit expression for $\eta_3$
takes the form:
\begin{eqnarray}
&&\eta_3=\frac{V^{(3)}}{\epsilon^{(3)}}=
\frac{V_1V_2V_3}{\tilde{\epsilon}_1+\tilde{\epsilon}_2+\tilde{\epsilon}_3+\tilde{U}_{12}+\tilde{U}_{13}+\tilde{U}_{23}}
\nonumber
\\
&&\times
\left[\frac{1}{\tilde{\epsilon}_1(\tilde{\epsilon}_1+\tilde{\epsilon}_2+\tilde{U}_{12})}
+\frac{1}{\tilde{\epsilon}_1(\tilde{\epsilon}_1+\tilde{\epsilon}_3+\tilde{U}_{13})}
\right.
\nonumber\\
&&\quad +\left.
\frac{1}{\tilde{\epsilon}_2(\tilde{\epsilon}_1+\tilde{\epsilon}_2+\tilde{U}_{12})}
+\frac{1}{\tilde{\epsilon}_2(\tilde{\epsilon}_2+\tilde{\epsilon}_3+\tilde{U}_{23})} \right.
\nonumber\\
&&\quad +\left.
\frac{1}{\tilde{\epsilon}_3(\tilde{\epsilon}_1+\tilde{\epsilon}_3+\tilde{U}_{13})}
+\frac{1}{\tilde{\epsilon}_3(\tilde{\epsilon}_2+\tilde{\epsilon}_3+\tilde{U}_{23})} \right].
\end{eqnarray}
This should be contrasted with the corresponding expression in the absence of diagonal interactions:
\begin{eqnarray}
\eta_3&=&\frac{V_1V_2V_3}{\epsilon_1+\epsilon_2+\epsilon_3}\nonumber\\
&\times& \left[\frac{1}{\epsilon_1(\epsilon_1+\epsilon_2)}+\frac{1}{\epsilon_1(\epsilon_1+\epsilon_3)}
+\frac{1}{\epsilon_2(\epsilon_1+\epsilon_2)}\right.
\nonumber
\\
&+&
\left.
\frac{1}{\epsilon_2(\epsilon_2+\epsilon_3)}+\frac{1}{\epsilon_3(\epsilon_1+\epsilon_3)}+\frac{1}{\epsilon_3(\epsilon_2+\epsilon_3)}\right]
\nonumber
\\
&=& \frac{V_1V_2V_3}{\epsilon_1\epsilon_2\epsilon_3}.
\end{eqnarray}

The conditions that the processes of $k$-th order lead to delocalization are obtained as straightforward generalizations of Eqs.~(\ref{res-2}) and (\ref{spec-diff-2}):
\be
\label{res-k}
|V^{(k)}| \gtrsim \Delta^{(N)}_{(k)}
\ee
and
\be
\label{spec-diff-k}
N \Delta^{(N)}_{(k)} \lesssim N^{1/2} V.
\ee
The level spacing $\Delta^{(N)}_{(k)}$  is of the order of
\be
\label{Delta-N-k}
\Delta^{(N)}_{(k)} \sim \frac{\left(\Delta^{(N)}\right)^k}{\epsilon^{k-1}} \sim \frac{\Delta^k}{N^{3k}\epsilon^{k-1}},
\ee
which is a generalization of Eq.~(\ref{Delta-N-2}).
The matrix element $V^{(k)}$ is estimated, in full analogy with Eq.~(\ref{V2-estimate}), as
\be
\label{Vk-estimate}
|V^{(k)}| \sim \frac{V^{k+1}}{\epsilon^k}.
\ee
Using Eqs.~(\ref{Delta-N-k}) and (\ref{Vk-estimate}), it is easy to see that the inequalities (\ref{spec-diff-k}) and  (\ref{Delta-N-k}) are compatible under the condition $N \gtrsim N_c^{(k)}$, where
\be
\label{Nck}
N_c^{(k)} \sim C_k g^{2k^2/(6k^2-1)},
\ee
with $C_k$ a $k$-dependent numerical coefficient.
For $k= 1$ and $k=2$ this result reduces to Eqs.~(\ref{E45}) and (\ref{Nc2}), respectively.

Taking the large-$k$ limit in Eq.~(\ref{Nck}), we conclude that the actual delocalization threshold is
\be
\label{Nc-quantum-dot}
N_c \sim g^{1/3}
\ee
up to subleading factors resulting from $C_k$ in Eq.~(\ref{Nck}).
This corresponds to the point where the number $p$ of resonances for a many-body basis states is of order unity, see Eq.~(\ref{p}). From this  point of view, the delocalization threshold in the considered model of an electronic quantum dot is similar to that in a finite system with power-law interactions \cite{gutman16}. In both cases, the appearance of just a few resonances is sufficient, by virtue of spectral diffusion, for many-body delocalization of the whole system. It is worth pointing out, however, that the way to delocalization in the present case is somewhat more complicated than in the model of Ref.~\cite{gutman16}. Specifically, in that model, the spectral diffusion was in fact a superdiffusion, which was so efficient that already the first-order resonances were sufficient to establish delocalization. On the other hand, in the present problem, such a treatment is sufficient only down to an intermediate scale (\ref{E45}) where the number $p$ of resonances is still parametrically large, $p \sim N^{1/2}$. In order to establish delocalization down to the scale (\ref{Nc-quantum-dot}), where $p\sim 1$, we had to invoke higher-order resonant processes.

\subsubsection{Comparison with tight-binding models on tree-like graphs: Logarithmic factor}
\label{s3A4}

The analysis presented in Sec.~\ref{s3A3} demonstrates that the actual delocalization border is given by Eq.~(\ref{Nc-quantum-dot}).
In Appendix~\ref{App:A} we give an alternative derivation of this result. The idea of this derivation is as follows.
We consider the regime $g^{1/3} < N < g^{2/5}$, so that the number of direct (first-order) resonances characterizing a typical many-body basis state is given by Eq.~(\ref{p}). Exploiting these resonances, one can proceed up to the generation $m$ given by Eq.~(\ref{m}), as explained in Sec.~(\ref{s3A1}). This yields a ``resonant subsystem'' formed by $m$ resonances, with the total number of many-body states estimated as ${\cal N}_r \sim \exp\{m\}$. All these basis states are well mixed, yielding states distributed roughly uniformly over them, with a level spacing $\sim  1/{\cal N}_r$. The resonant subsystem now serves as a kind of ``bath'' that assists decay processes for other electrons (similar to the spin bath model \cite{prokofiev00}). Of course, we deal with a finite system, so that the ``bath'' has in fact a discrete spectrum. What helps, however, is that the level spacing is exponentially small. As a result, already for a logarithmically large number of resonances, this level spacing becomes smaller than the characteristic energy scales of the electron decay process, so that the resonant subsystem can serve essentially as a continuous bath. Specifically, a detailed analysis presented in Appendix~\ref{App:A} leads to Eq.~(\ref{Nc-log}) for the energy above which the whole system gets delocalized via this mechanism. This result agrees with Eq.~(\ref{Nc-quantum-dot}), up to a logarithmic factor $(\ln g)^{1/4}$.

We thus have two different arguments that lead us to the conclusion that the delocalization border $N_c$ for the electronic quantum dot problem scales with $g$ as
  \be
 \label{Nc-log-mu}
 N_c \sim g^{1/3} (\ln g)^{\mu}.
 \ee
 The analysis of Appendix~\ref{App:A} provides us with the upper bound for the shifting down of the delocalization
 transition point: $\mu \le 1/4$.
 On the other hand, a logarithmic enhancement ($\mu<0$) of the delocalization is known to take place for the Anderson transition in tight-binding models on the Bethe lattice and related tree-like graphs, such as random regular graphs (RRG). In fact, the Fock-space structure of the considered model is similar to an RRG with an average number of resonantly coupled  neighbors for each vertex equal to $p = N^3/g$. The position of the Anderson transition for such an RRG model will be
 \be
 \label{pc}
 p_c \sim 1/\ln g,
 \ee
 yielding
 \be
 \label{Nc-RRG}
 N_c^{\rm RRG} \sim \left(\frac{g}{\ln g}\right)^{1/3},
 \ee
 which has a form of  Eq.~(\ref{Nc-RRG}) with $\mu = -1/3$.
 This value of $\mu$ would correspond to the limit of $U\to \infty$,
 where the Bethe-lattice structure of the matrix elements (\ref{Vk-exact}) is restored.
 Therefore, we conclude that for $U\sim V$ the delocalization threshold in a quantum dot is characterized by
 \begin{equation}
 -\frac{1}{3}\leq \mu \leq \frac{1}{4}.
 \label{estimate-mu}
 \end{equation}

At this point, it is worth discussing whether the mechanism that leads to the  logarithmic enhancement of the position of the delocalization threshold in the RRG model,  Eq.~(\ref{Nc-RRG}), is operative in the many-body problem. The emergence of the logarithm in the Bethe-lattice and RRG problems can be seen by starting from the localized phase and inspecting the probabilities to have resonances of order $k \ge 2$.  For $k = 2$, the effective matrix element is then
 \be
 \label{V2-RRG}
 |V^{(2)}|  \sim  \frac{V^2}{\epsilon},
 \ee
 where $\epsilon$ is the energy of the intermediate state. The level spacing $\Delta_{(2)}$ of final states is given, under the condition that the energy of the intermediate state is between $\epsilon$ and $2\epsilon$, by
\be
\label{Delta-2-RRG}
\Delta_{(2)} \sim \frac{V^2}{p^2\epsilon}.
\ee
Using   Eqs.~(\ref{V2-RRG}) and (\ref{Delta-2-RRG}), one finds that the probability to have a resonance is $\sim  |V^{(2)}| / \Delta_{(2)}  \sim p^2$, independently of $\epsilon$. Summing over the windows $(\epsilon, 2\epsilon)$ of the intermediate-state energies amounts, in the continuous version, to evaluation of the integral $\int d\epsilon / |\epsilon| $, yielding a logarithmic factor (translating to $\ln g$ in our notation). Similarly, in the third order an integral of the type $\int d\epsilon_1 d\epsilon_2 / |\epsilon_1 \epsilon_2|$ emerges, and so on. As a result, the probability to have a resonance in $k$-th order scales as $p^k (\ln g)^{k-1}$, thus yielding Eq.~(\ref{pc}).

Extending this analysis to the quantum-dot many-body problem with $U\sim V$, we observe the following difference (see Appendix~\ref{App:1}). The matrix element  $V^{(2)}$ is given in the case of a quantum dot  by Eq.~(\ref{V2-estimate}), i.e., it is suppressed as compared to the case of a Bethe-lattice (or RRG),
Eq.~(\ref{V2-RRG}), by the additional small factor $V/\epsilon$ resulting from  a partial cancelation of two contributions. As a result, the probability to have a resonance in the second order will now include, in place of $\int d\epsilon/|\epsilon|$ appearing in Bethe-lattice-like models, an integral $V \int d\epsilon/\epsilon^2$. This integral is not logarithmic anymore but rather is determined by its lower limit, which is $\epsilon \sim \Delta/N^3$ (for $\Delta/N^3 \lesssim V$), thus yielding simply a factor $p = N^3 /g \lesssim 1$. The same happens at higher orders. As an example, to third order,  we obtain, instead of the Bethe-lattice integral
$\int d\epsilon_1 d\epsilon_2 / |\epsilon_1 \epsilon_2|$, an integral of the type
$V \int d\epsilon d\epsilon' / |\epsilon \epsilon' (\epsilon+\epsilon')|$, which is again not logarithmic in $g$.
As a consequence, in the many-body problem the probability to find a resonance of $k$-th order
scales with $g$ as $p^{k+1}\propto 1/g^{k+1}$.
This suggests that the resonances start to proliferate at
$p_c \sim 1$,
instead of Eq.~(\ref{pc}) valid for a tree-like model.

However, this estimate ignores the non-parametric $k$-dependence of the above integrals [cf. factors $C_k$ in Eq.~(\ref{Nck})].
In particular, at $k$-th order the accumulation of elementary energy shifts due to the diagonal elements is expected to
lead to the appearance of a typical overall shift $\sqrt{k} U$, similar to the spectral-diffusion picture.
Then, for $U\sim V$, there appears the energy window $V<\epsilon<\sqrt{k}V$ where a logarithmic factor $\ln k$ can be accumulated,
replacing the $\ln g$-factor in the probability to find a resonance, see Appendix~\ref{App:1}.
Taking the maximum value $k\sim N_c$, one would recover the $\ln g$-factor in the critical value of $p$.
This would, in turn, lead to Eq.~(\ref{Nc-log-mu}) with $-1/3\leq \mu<0$.

A more detailed study of the logarithmic factor in the position of the delocalization threshold is relegated to future work.
For the purposes of the present paper, we only establish the bounds on $\mu$, see Eq.~(\ref{estimate-mu}) above.

 \subsection{Spin quantum dot}
\label{s3B}

We turn now to the spin quantum dot model defined in Sec.~\ref{s2C}. Its analysis proceeds in the same way as for the electronic quantum dot in Sec.~\ref{s3A}.  The role of many-body basis states is played by the eigenstates of all $S_k^z$ operators. The first term in the Hamiltonian (\ref{H-spin-quantum-dot}) induces transitions between them. (Each of such transitions corresponds to flipping two spins or a single spin.)
The dimensionless interaction coupling $\alpha$ plays the same role as $1/g$ in the formulas of Sec.~\ref{s3A}. The level spacing of all states resonantly coupled to a given basis state is now
\begin{equation}
\Delta^{(N)}\sim   \frac{\Delta}{N^2},
\label{DeltaN-spin}
\end{equation}
which is larger by a factor of $N$ than that in the quantum dot model of Sec.~\ref{s3B}, see Eq.~(\ref{DeltaNtyp}). This is the only essential change, which correspondingly modifies all the results.
Discarding the spectral diffusion, one finds, in analogy with Eq.~(\ref{T1log}), the ``conservative estimate''  for the transition point:
\begin{equation}
\label{Nc-spin-nsd}
N_c^{\rm nsd} \sim \frac{1}{\alpha\ln(1/\alpha)}.
\end{equation}
Up to a logarithmic factor, Eq.~(\ref{Nc-spin-nsd}) is obtained by comparing the matrix element (\ref{DeltaN-spin}) with the level spacing $N\Delta^{(N)} \sim \Delta/N$ of states corresponding to the first-order decay of a given spin. The total number of first-order resonances (due to processes with two spins flipped) in a given many-body basis state is
\be
\label{p-spin}
p \sim  \frac{V}{\Delta^{(N)}} \sim \alpha N^2,
\ee
which is a counterpart of Eq.~(\ref{p}). The critical point is determined by the condition $p_c\sim 1$, yielding
\be
\label{Nc-spin}
N_c \sim \alpha^{-1/2},
\ee
up to a possible logarithmic factor $\ln^\mu|\alpha|$ (see discussion in Sec.~\ref{s3A4}).
The delocalization for $N \gtrsim \alpha^{-2/3}$ is obtained by considering the first-order processes and including into consideration the spectral diffusion, in full analogy with Sec.~\ref{s3A1}. In this range, the spin relaxation rate is given by the conventional golden-rule formula,
\be
\frac{1}{\tau} = \frac{1}{N}\Gamma_{\rm MB} \sim \frac{V^2}{\Delta/N} \sim \alpha^2 N \Delta,
\label{Gamma-MB-spin}
\ee
in analogy with Eqs.~(\ref{GR1}), (\ref{GR2}). On the other hand, to demonstrate delocalization for $N$ below $\alpha^{-2/3}$ [down to the critical value (\ref{Nc-spin})], one should invoke higher-order processes, in analogy with Secs.~\ref{s3A3} and \ref{s3A4}.

\section{Enhancement of delocalization by spectral diffusion in spatially extended systems }
\label{s4}

In Sec.~\ref{s3}, we have analyzed the scaling of the many-body delocalization threshold in two models of a quantum dot. In the present section, we expand the analysis to the case of spatially extended systems. We first consider the spin system of Sec.~\ref{s2D} and then generalize the results to the electronic model with spatially localized single-particle states, Sec.~\ref{s2B}.

\subsection{Spin system}
\label{s4A}

Here we analyze the spin system defined by the Hamiltonian (\ref{H-spin-extended}). It can be viewed as a system of coupled spin quantum dots considered in Sec.~\ref{s3B}, with the interaction of spins in adjacent dots being of the same order as within a dot. We argue now that the scaling of the delocalization transition is given by the same formula as for the spin quantum dot,
Eq.~(\ref{Nc-spin}), where $N$ is the number of spins in each dot. Let us first show that for $N$ exceeding this value the system is definitely delocalized. We present two somewhat different (though, of course, related) ways of reasoning.

Consider a typical basis state of the whole system (which is a product of the basis states of all dots). The number $p$ of resonant spin pairs in each dot, or in a pair of adjacent dots, is given by Eq.~(\ref{p-spin}). Let us flip one such resonant pair. Because of spectral diffusion, this will create $\sim p$ new resonant pairs in adjacent dots (or pairs of dots), thus inducing $\sim p$ new transitions. We can again choose any of them and apply the same reasoning, with an excitation moving to  a new dot on each step. As a result, a structure characteristic of the Bethe lattice emerges, in a certain analogy with the processes of a quasiparticle decay considered in Refs.~\cite{gornyi05} (where they were called ``ballistic'') and \cite{ros15} (where the term ``necklace diagrams'' was used).
This Bethe-lattice-like structure is characterized by the number $p$ of states well coupled at each step. This ensures that for $p\gtrsim 1$ the system should be in the delocalized phase.

The second version of this argumentation is in the spirit of the approach used for the analysis of systems with power-law interaction, see Ref.~\cite{gutman16}  and references therein. Let us assume that the system is in the localized phase. In each quantum dot we have $\sim p$ resonant spin pairs. Two flip-flop states of such a pair can be considered as forming a pseudospin.  Restricting our attention to a subspace formed by such pseudospins, we get an effective Hamiltonian of the pseudospin subsystem (``resonant subsystem''), which has the same form as Eq.~(\ref{H-spin-extended}) but with the number of pseudospins per dot being $p$ (instead of $N$) and with the ratio of the matrix element to the typical Zeeman field being unity (instead of $\alpha$). This system clearly undergoes a many-body delocalization transition at some $p_c\sim 1$. Thus, above this $p_c$ the assumption about the localization of the original system was incorrect: the localized phase is unstable.

We have thus shown that the delocalization extends at least down to the value of $N$ given by  Eq.~(\ref{Nc-spin}).
As in the spin quantum dot, the spin relaxation rate for $N \gtrsim \alpha^{-2/3}$ is given by the conventional golden-rule formula (\ref{Gamma-MB-spin}). The relaxation rate closer to the transition, i.e., for $ N_c < N \ll \alpha^{-2/3}$ remains to be determined.

It is worth pointing out an interesting property of the localized phase (which is in fact a general feature of models of many-body localization with spatially localized single-particle states). In the localized phase, $N \ll N_c$, the average number of resonances per dot is much smaller than unity, $p \sim \alpha N^2 \sim (N / N_c)^2 \ll 1$. This means that in most of the quantum dots forming the system there are no resonances at all, while in a fraction $p$ of them there is a single resonance. If we consider a large but finite system with $M$ dots, there will be $pM$ dots with a resonance. A standard characteristic of the degree of localization of an eigenstate $\Psi$ is its  inverse participation ratio (IPR)
\be
\label{IPR}
P_2 = \sum_i |\langle \psi_i  | \Psi \rangle |^4,
\ee
where $\psi_i$ are the basis states (which are eigenstates in the limit of complete localization). For noninteracting systems the localized phase is characterized by $P_2 \sim 1$, while the delocalized phase by $P_2 \propto 1/ {\cal N}$, where ${\cal N}$ is the size of the Hilbert space. For our many-body system we have ${\cal N} = 2^{NM}$. Further, according to the above discussion, the IPR in the localized phase is estimated as $P_2 \sim 1/2^{cpNM} = {\cal N}^{cp}$, with $c \sim 1$, where we used the fact that a quantum dot where there are no resonances is roughly in one of its basis states, while a dot with
a resonance is approximately equally spread over its two basis states. This yields
\be
P_2 \sim 1/ {\cal N}^{\tau}, \qquad {\rm with} \qquad  \tau \sim p,
\ee
i.e., the IPR in the many-body localized phase scales as a fractal power of the total number of many-body states ${\cal N}$. While for single-particle system such a fractality is a very nontrivial property that emerges only at the Anderson-transition point, in the case of a many-body system it represents, for a rather simple reason, a general property of the localized phase. This behavior of IPR is closely related to the volume-law scaling of the long-time saturation value of the entanglement entropy for the case when the initial state is a basis state \cite{bardarson12,serbyn13}.

\subsection{Electron system}
\label{s4B}

In view of the close connection between the spatially extended spin system analyzed in Sec.~\ref{s4A} and the electronic system defined in Sec.~\ref{s2B}, the result for the delocalization threshold in the former model, Eq.~(\ref{Nc-spin}), applies to the latter model as well. Translating the value of $N$ into a temperature according to the ``dictionary" presented in Sec.~\ref{s2D}, we obtain the critical temperature
 \be
\label{Tc12}
T_c \sim \frac{\Delta_\xi }{\alpha^{1/2}},
\ee
up to a possible logarithmic factor $\ln^\mu(1/\alpha)$ with $-1/2\leq \mu\leq 0.$
We note that the upper bound for $\mu$ is zero, since the level spacing in the resonant network (Appendix~\ref{App:A})
in an extended system goes to zero.

The result (\ref{Tc12}) is parametrically lower than the result (\ref{Tc-ext-nsd}) of Refs.~\cite{gornyi05,basko06,ros15}. At this temperature both energy and charge get delocalized. The energy delocalization is fully analogous to that in the spin model discussed in Sec.~\ref{s4A}. The existence of charge excitations is what distinguishes the electronic model from the spin one. In the electronic model, each single electron state can form a ``spin'' with a few partner states (separated by energy $\sim \Delta_\xi$) in the same or adjacent quantum dot, which leads to charge transport. It is expected that the spin and charge delocalization do not only have the same scaling (\ref{Tc12}) but actually happen at the same point. Indeed,  energy transport implies the existence of a ``bath'' to enable charge transport as well.

\begin{figure*}
  \centering
  \includegraphics[width=0.95\textwidth]{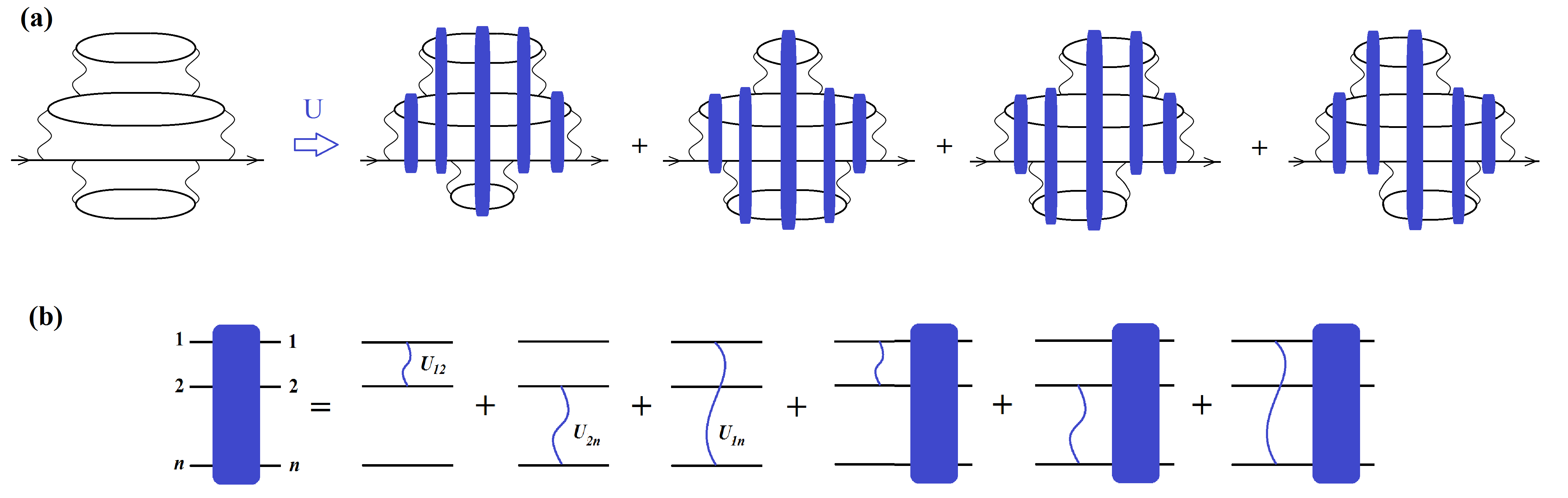}
  \caption{(a) Self-energy diagrams for single-particle decay: the dressing of a skeleton diagram by diagonal
  interactions $U$
   produces a factorial number of nonequivalent Feynman diagrams. Blue blocks represent full ladders of all possible diagonal interactions between $n$ particles, as shown in panel (b).}
  \label{Fig:self}
\end{figure*}

The result (\ref{Gamma-MB-spin}) for the relaxation rate translates into the golden-rule formula
\be
\label{tau-electron-ext}
1/\tau \sim \alpha^2 T, \qquad T \gtrsim \Delta/\alpha^{2/3}.
\ee
This can be used to estimate the conductivity. A convenient quantity is the dimensionless conductance of a piece of the system of linear size $\xi$:
\be
\label{power-law-hopping}
g_\xi \sim \frac{1}{\tau\Delta} \sim \alpha^2 \frac{T}{\Delta},  \qquad T \gtrsim \Delta/\alpha^{2/3}.
\ee
This transport regime was termed ``power-law hopping'' in Ref.~\cite{gornyi05};
we now see that its range of validity extends further down to parametrically lower temperatures in comparison with what was found in that work.
The behavior of the relaxation rate and of the conductivity closer to the transition point, i.e., at $T_c < T \ll \Delta/\alpha^{2/3}$ remains to be determined.

\subsection{Feynman diagrams and factorials}
\label{s4C}

Let us now discuss how the above results could be obtained in the Feynman diagrammatic approach employed in Refs.~\cite{gornyi05,basko06,ros15}, where a single-particle decay was considered.
 A typical self-energy diagram describing such a decay is shown in Fig.~\ref{Fig:self}.
 In the absence of diagonal matrix elements $U$, the time-ordering of the off-diagonal interactions $V$ that produces
 distinct paths in Fock space (Fig.~\ref{Fig:Fock}) within the Schr\"odinger perturbation theory, is redundant.
 All such paths are contained in a single Feynman diagram (skeleton diagram in Fig.~\ref{Fig:self}) \cite{silvestrov97,silvestrov98,basko06,ros15,gornyi16}.

 Let us first show how the threshold (\ref{T1log}) for the quantum dot is obtained within an estimate of such skeleton diagrams in the spirit of Refs.~\cite{gornyi05,ros15}. We consider processes of order $n$ corresponding to transitions between the original single-electron excitation and the final state with $2k+1$ quasiparticles $\beta_j$ ($k+1$ electrons and $k$ holes).   The final state corresponds to a central cross-section of the skeleton diagram of the type shown in Fig.~\ref{Fig:self}a; see also Fig. 1b or Ref.~\cite{gornyi05}.

 The matrix element of a transition from an initial to the final state is given by \cite{gornyi05}
 \be
 V^{(k)} = \sum_{\rm diagrams} \sum_{\gamma_1 \ldots \gamma_{k-1}} V_1 \prod_{i=1}^{k-1} \frac{V_{i+1}}{E_i - \epsilon_{\gamma_i}}\:,
 \label{Vn}
 \ee
 where the matrix element $V_i$ corresponds to the $i$-th interaction line, $\gamma_i$ are the virtual quasiparticles corresponding to the internal quasiparticle lines in the diagram, and $E_i$ are the associated energy variables that can be expressed as linear combinations of the energies of the final-state quasiparticles $\epsilon_{\beta_j}$ using the energy conservation. Note the absence of interaction matrix elements in the denominator of Eq.~(\ref{Vn}), which is in contrast to the structure of
 the composite matrix element (\ref{Vk-exact}).

 The number of topologically different diagrams, $D^{(k)}$, satisfies the recurrency relation \cite{gornyi05}
 \be
 D^{(k)} = \sum_{k_1+k_2+k_3=k-1; \ k_{1,2,3}\ge 0}  D^{(k_1)} D^{(k_2)} D^{(k_3)}\:,
 \ee
 with the initial condition $ D^{(0)}=1$. The solution of this equation increases with $k$ as $ D^{(k)} \sim a^k$ with $a \sim 1$ (the numerical value of $a$ was determined in Ref.~\cite{ros15}). We will not be interested in such factors here (since they will only produce factors of order unity for the localization threshold), so that we can replace $D^{(n)}$ by unity.
 The summation over $\gamma_i$ can be replaced by taking the ``optimal'' $\gamma_i$, which renders the corresponding energy denominator  in Eq.~(\ref{Vn}) to be of order $\Delta$. For a given set of final-state quasiparticles $\beta_j$ there is $M^{(k)} = k! (k+1)! \sim k^{2k}$ permutations of the corresponding lines in the diagram. The dominant contribution will be given by the optimal permutation~\cite{ros15}, thus yielding
 \be
  V^{(k)} \sim M^{(k)} V^{k+1} \Delta^{-k} \sim \Delta \left(\frac{k^2}{g}\right)^k .
  \label{Vk}
  \ee
The level spacing of final states of the generation $k$ is estimated as
  \be
  \Delta_{(k)} \sim \Delta \left(\frac{k}{N}\right)^{2k}.
  \ee
Therefore, we find for the parameter
$\eta_k = V^{(k)}/\Delta_{(k)} $
controlling the hybridization at $k$-th order:
  \be
  \eta_k \sim  \left(\frac{N^2}{g}\right)^k .
  \label{eta-k}
  \ee
This yields $N_c \sim g^{1/2}$ for the position of the localization transition. A more accurate analysis of the sums over intermediate states leads to the emergence of an additional factor $ \sim (\ln N)^{k-1} $ in Eq.~(\ref{Vk}), thus yielding $N_c \sim (g/\ln g)^{1/2}$,  which is the threshold (\ref{T1log}).

As emphasized above, this estimate used in a very essential way the fact that  a single skeleton diagram represents all processes with different time orderings of the interaction lines.  However, this is no longer the case when the skeleton diagram is dressed
 by diagonal interactions. Indeed, as shown in Fig.~\ref{Fig:self}a, such a dressing generates a factorial number of
 nonequivalent \textit{Feynman} diagrams, when the diagonal matrix elements, combined in multi-leg ladders as illustrated in Fig.~\ref{Fig:self}b, are inserted in all possible ways into the skeleton diagram. These ladders introduce the diagonal matrix elements $U$ into the energy denominators in the expressions for the diagrams, leading to the structures of the type $\sum_{ij} \tilde{U}_{ij}$ that
appear in the composite matrix element $V^{(k)}$ in Eq.~(\ref{Vk-exact}) (similarly to the appearance of the
interaction in the denominator of an exciton propagator). The spectral diffusion is thus encoded in the energy shifts in these denominators which are accumulated with increasing order of the diagram and lead to the factorial growth of the number of distinct Feynman diagrams.

Thus, in terms of a single-electron excitation decay, the enhancement of delocalization (in comparison with the results found for the extended system in Refs.~\cite{gornyi05,basko06} and for a quantum dot in Ref.~\cite{gornyi16}) is due to the $k!$ factor in the number of processes at $k$-th order (due to permutations of $k$ elementary processes mediated by the off-diagonal interaction).
If one discards spectral diffusion, these $k!$ contributions combine to a single one, yielding the results of the above works \cite{note-diag-BAA}.
We discuss now the effect of the spectral diffusion on the localization threshold, first for a quantum-dot model and then for an extended system.

In the case of a quantum dot, including spectral diffusion essentially yields an additional factor $k!$ (given by the number of diagrams generated by a single skeleton diagram, see Fig.~\ref{Fig:self}a). We thus find, instead of Eq.~(\ref{eta-k}),
\be
\eta_k\sim \left(\frac{N^2}{g}\right)^k k! \sim \left(\frac{N^2 k}{g}\right)^k.
\ee
Taking in this formula $k$ to be the maximum possible (i.e., of the order of  $N$, the number of active single-particle states), we get the following estimate for the effective coupling controlling the perturbative expansion of the self-energy (cf. Ref.~\cite{silvestrov98}):
\be
\eta_k \ \stackrel{k \sim N}{\rightarrow} \   \left(\frac{N^3}{g}\right)^N,
\label{eta-k-factorial}
\ee
Equating $\eta_{k \sim N}$ to unity,  we recover the result of Sec.~\ref{s3A} for the localization threshold, see Eq.~(\ref{Nc-log-mu}).

A word of caution is in order at this point. What we have shown in the preceding part of this subsection is that (i) when spectral diffusion is included, a single skeleton Feynman diagram generates $\sim k!$ diagrams, and (ii) when all of them are considered as independent, we reproduce the result (\ref{Nc-log-mu}) for the localization threshold in a quantum dot. However, there are clearly certain remaining correlations between these $k!$ diagrams, and, in order to justify rigorously the appearance of the whole $k!$ factor in Eq.~(\ref{eta-k-factorial}), one would have to prove that these correlations are not essential. Thus, the above arguments based on Feynman diagrammatics make the result (\ref{Nc-log-mu}) plausible but, strictly speaking, do not prove it. We have shown, however in Sec.~\ref{s3A}, by using  an alternative approach based on the perturbation theory for Fock-space states, that Eq.~(\ref{Nc-log-mu}) is indeed correct.

For an extended system, including the total factor $k!$ into the number of diagrams (as was done in Ref.~\cite{gornyi04}) would lead to the absence of the genuine many-body localization in the thermodynamic limit \cite{remark1}. However, for a short-range interaction there is no contribution to spectral diffusion due to interactions between remote ``quantum dots'' (localization volumes). Thus, only the factorial factors originating because of the permutation of processes within a given quantum dot should survive. Writing $k = M N$, where $M$ is the number of quantum dots involved, we see that out of $k! \sim k^k$ only the factor   $(N!)^M \sim N^{NM} = N^k$ survives. This will exactly transform the result of Refs.~\cite{gornyi05,basko06} into Eq.~(\ref{Tc12}):
\begin{equation}
\eta_k\sim \left(\frac{V}{\Delta_3(T)}\right)^k (N!)^M \sim \left(\frac{\alpha T N}{\Delta_\xi}\right)^k  \sim
  \left(\alpha \frac{T^2}{\Delta_\xi^2}\right)^k.
\end{equation}

In full analogy with the quantum-dot case, these Feynman-diagrammatics arguments make plausible the emergence of the factor $N^k$ in $\eta_k$ because of spectral diffusion---which leads to Eq.~(\ref{Tc12}) for the delocalization threshold---but, strictly speaking, do not prove it. We know, however, on the basis of the arguments given in Sec.~\ref{s4}, that Eq.~(\ref{Tc12}) is indeed correct.

Within this picture, delocalization in extended systems occurs by means of consecutive delocalization in constituent
 ``quantum dots''. The corresponding paths in both Fock and real spaces can be viewed as ``ballistic'' (in the sense of Ref.~\cite{gornyi05} or, equivalently, in the sense of necklace diagrams from Ref.~\cite{ros15}) after
``coarse-graining'' in which each step corresponds to delocalization of one quantum dot. At this stage, we can not fully exclude a possibility that spectral diffusion might potentially further lower the threshold temperature when included into other (``non-ballistic'') processes.
We relegate this issue to future work.

\section{Summary and discussion}
 \label{s6}

To summarize, we have shown that taking into account the spectral diffusion, which originates from diagonal matrix elements of the interaction, parametrically enhances the delocalization threshold in many-body systems. This happens both in quantum-dot models and in spatially extended models with localized single-particle states. In particular, in an electronic system of the latter type, the critical temperature $T_c$ of the delocalization transition scales with the interaction strength  $\alpha \ll 1$ according to Eq.~(\ref{Tc12}), which is parametrically lower than the previous result (\ref{Tc-ext-nsd}) of Refs.~\cite{gornyi05,basko06,ros15}.

Before closing, let us discuss some further implications of our work as well as directions for future research.

\begin{description}

\item{(i)}  Our analysis shows that the localization transitions in quantum-dot many-body problems have a lot of similarity to those in tight-binding models on tree-like graphs, such as the RRG \cite{tikhonov16} and sparse random-matrix \cite{srm} models. The essential difference  is in the fact that for the effective tight-binding model of a quantum dot there are small loops [e.g., the processes $(\alpha,\beta \to \gamma,\delta)$ and $(\alpha',\beta' \to \gamma',\delta')$ performed in opposite order results in the same state] and that the corresponding amplitudes partially cancel. This cancelation
affects the analysis of logarithmic corrections to the scaling of the delocalization threshold, see Sec.~\ref{s3A4}. At the present stage, we are only able to give boundaries, Eq.~(\ref{estimate-mu}), for the power of the logarithm $\mu$ in the scaling law, Eq.~(\ref{Nc-log-mu}). An accurate determination of $\mu$ remains an important research prospect.

Further, it would be interesting to study whether (and, if yes, how) the above  difference between the quantum-dot problem and the RRG affects the critical behavior at the transition. In particular, one can study the scaling of the wave function statistics (inverse participation ratios), and the level statistics in the quantum-dot problem, at criticality and near the transition.

\item{(ii)} The problem of exact determination of the power of the logarithmic factor in the scaling of the threshold applies also to spatially extended models with localized single-particle states, see a comment below Eq.~(\ref{Tc12}).  Clearly, the scaling of observables at and near the transition in this class of many-body systems is of great interest as well. Three of us have argued in Ref.~\cite{gornyi05}, by using a relation of the delocalization of a quasiparticle excitation by ``ballistic'' paths to the Bethe lattice, that the quasiparticle relaxation rate and the conductivity scale near critical point as
$$\sigma \propto \exp\left[-{\rm const}\times(T-T_c)^{-1/2}\right].$$
  While the spectral diffusion modifies $T_c$, it does not seem to affect the argument in favor of this scaling. Indeed, for a spatially extended system, the $k$-dependence of the $k$-th order effective matrix element for a quasiparticle decay remains for large $k$ the same as in the Bethe-lattice problem also with spectral diffusion taken into account.
  Clearly, the above argument is not rigorous since it discards other contributions to the quasiparticle decay. A rigorous derivation of the critical behavior in a many-body problem remains
an outstanding challenge for future research.

It is worth mentioning that an experimental evidence in favor of an exponential vanishing of the conductivity at the threshold was recently obtained in Ref.~\cite{ovadia15} in a 2D system on the localized side of the superconductor-insulator transition.

\item{(iii)} Several numerical studies of 1D systems have reported a subdiffusive transport on the delocalized side of the transition \cite{agarwal15,barlev15,gopalakrishnan15,lerose15}. The existence of such a phase has been ascribed to rare (Griffiths-type) events \cite{agarwal15,vosk15,ACP15,knap15,agarwal16}. An important question is whether this subdiffusive phase really exists in the thermodynamic limit.
Recent work \cite{bera16} observed a substantial flow of the corresponding fractal dimension with the system size (in a certain analogy with Ref.~\cite{tikhonov16} for RRG), which suggests that the thermodynamic-limit behavior might differ essentially from the one observed numerically in relatively small systems.

Signatures of many-body localization were also studied in systems without quenched disorder~\cite{kagan84,grover14a,schiulaz14a,yao14a,deroeck14,papic15,mueller16}.
A distinction between the models with and without quenched disorder in the problem of many-body localization was emphasized in Ref.~\cite{deroeck14}.
In a related line of reasoning, it was conjectured in Refs.~\cite{ros15,mueller16} that the rare events of the ``hot bubble'' type may destroy the many-body localized phase. It is interesting to note that the rare events were argued to lead both to suppression and to enhancement of localization, see discussion in Refs.~\cite{agarwal16} and \cite{deroeck16}.

It remains an open issue to understand whether the rare-region effects of both types can be captured within the scheme developed in the present work, with the spectral diffusion included.

\item{(iv)}
A verification of the analytical predictions for the scaling of the many-body localization threshold by means of numerical simulations would be of great interest. Several papers addressed this scaling numerically in the framework of the electronic quantum-dot problem fifteen to twenty years ago \cite{jacquod97,georgeot97,shepelyansky01,leyronas00,rivas02,jacquod02} with
seemingly contradictory conclusions.
Specifically, Refs.~\cite{jacquod97,georgeot97,shepelyansky01,jacquod02} concluded that the localization threshold scales as $N_c \sim g^{1/3}$, which is in agreement with our result  Eq.~(\ref{Nc-log-mu}) (possibly up to a logarithmic factor). On the other hand, Ref.~\cite{leyronas00}
came to the conclusion that the scaling is of the form $N_c \sim (g/ \ln g)^{1/2}$, which is the result (\ref{T1log}) obtained within the approximation~\cite{gornyi16} that discards the spectral diffusion. It should be emphasized, however, that Ref.~\cite{leyronas00} considered a model of the type (\ref{ham}) which included only interaction terms with four different indices $\alpha$, $\beta$, $\gamma$, and $\delta$. This absence of diagonal terms in the Hamiltonian of Ref.~\cite{leyronas00} reduced the effect of the spectral diffusion, which may explain why the threshold found in Ref.~\cite{leyronas00} corresponds to Eq.~(\ref{T1log}) that neglects the spectral diffusion.
Clearly, exploring somewhat larger systems would be beneficial for a clear identification of the power-law exponent. Probably, it would be easier to distinguish between the different types of scaling in the spin quantum dot model formulated in Sec.~\ref{s2C}.

In the case of spatially extended models, the phase diagrams in the interaction-disorder plane have been established numerically in Refs. \cite{barlev15} and \cite{luitz15}. While a determination of the scaling of the critical disorder with the interaction strength has not been attempted in these works, it appears to be feasible (although clearly not simple). As an approach complementary to exact diagonalization, a numerical evaluation of a sum over forward-approximation paths \cite{pietracaprina16} might be used for this purpose.

The authors of Ref.~\cite{cuevas12} analyzed specifically the role of diagonal matrix elements in many-body delocalization in the framework of a spin model on a random regular graph. Numerical calculations performed in Ref.~\cite{cuevas12} are in a qualitative agreement with the spectral-diffusion picture:  the additional diagonal terms substantially enhance delocalization.

\item{(v)} It is worth emphasizing that, when considering spatially extended systems in this work, we assumed (following Refs.~\cite{gornyi05,basko06,ros15}) that the single-particle localization length may be considered as an energy-independent quantity. This approximation is clearly justified if one considers a tight-binding lattice model with energy somewhere in the middle part of the band. On the other hand, for systems defined in a spatial continuum, one encounters a situation in which there are states with an arbitrarily large single-particle localization length (see Refs.~\cite{nandkishore14,potter14,li15} for the discussion of many-body localization in systems with an energy-dependent single-particle localization length). Of course, these are highly excited states whose thermal occupation is exponentially low, so that their effect is not clear a priori and has to be carefully investigated. An analysis shows~\cite{continuum}  that scattering between thermal and high-energy states eliminates the finite-$T$ many-body localization transition, rendering many-body states delocalized in the thermodynamic limit at arbitrarily low (but still nonzero) temperature. In such a situation, the critical temperature identified in the present work will become a crossover temperature below which the conductivity tends to zero in a faster-than-activation fashion.

\end{description}

\section*{Acknowledgements}

We would like to thank  M.V.~Feigel'man, M.~M\"uller, A.~Scardicchio, P.G.~Silvestrov, D.L.~Shepelyansky, and K.S.~Tikhonov for discussions.
We are particularly grateful to I.V.~Protopopov for collaboration on the early stage of this work
and instructive discussions
A.L.B. is grateful to Karlsruhe Institute of Technology for hospitality extended to him during his visit in May 2016.
This work was supported by Russian Science Foundation under
Grant No.\ 14-42-00044 (I.V.G. and A.D.M.) and by the National Science Foundation under Grant CHE-1462075
(A.L.B.).

\appendix

\section{Distribution function of resonant couplings}
\label{App:1}

In this Appendix we calculate the distribution function of the effective coupling constant
for second-order processes (Fig.~\ref{Fig:2res}) and estimate the number of $k=2$
resonances in a quantum dot for given values of $N$ and $g$.

For given values of the energy mismatches $\epsilon$, $\epsilon^\prime$, off-diagonal interactions $V$, $V^\prime$, and
diagonal interaction $U$ between the two parallel processes, the coupling constant $\eta_2$ is given by the
ratio of the composite matrix element $V^{(2)}$, Eq.~(\ref{V2}),
and the total energy mismatch $\epsilon^{(2)}$, Eq.~(\ref{epsilon2}).
We define the function
\begin{eqnarray}
\mathcal{F}_2(x)&=&\int_x^\infty d\eta \int d\epsilon\ p_E(\epsilon) \int d\epsilon^\prime\ p_E(\epsilon^\prime)\nonumber\\
&\times& \int dU \mathcal{P}_u(U)\int dV \mathcal{P}_v(V)\int dV^\prime \mathcal{P}_v(V^\prime)\nonumber \\
&\times&
\delta\left(\eta-\frac{V V^\prime (\epsilon+\epsilon^\prime)}{\epsilon\epsilon^\prime(\epsilon+\epsilon^\prime+U)}\right),
\label{F2def}
\end{eqnarray}
which gives the number of $k=2$ processes with the coupling constant exceeding $x$.
The energy distribution function $p_E(\epsilon)$ describes the thermal distribution of four single-particle energies
characterized by $T\sim \Delta N$;
for simplicity, we approximate it by the homogeneous box distribution (in what follows, we omit the numerical prefactors):
\begin{equation}
p_E(x)\sim \frac{N^3}{\Delta}\theta(N\Delta-|x|).
\end{equation}
The distribution of the interaction matrix elements is Gaussian:
\begin{eqnarray}
\mathcal{P}_u(x)&\sim& \frac{1}{U_0} \exp\left(-\frac{x^2}{U_0^2}\right), \label{PU}\\
\mathcal{P}_v(x)&\sim& \frac{1}{V_0} \exp\left(-\frac{x^2}{V_0^2}\right). \label{PV}
\end{eqnarray}
In the problem under consideration, $U_0\sim V_0\sim \Delta/g$, but it is instructive to
keep the typical values of the diagonal and off-diagonal matrix elements nonequal.
Writing the delta-function in Eq.~(\ref{F2def}) as an integral over the auxiliary variable,
we integrate out $V$ and $V^\prime$, which yields
\begin{eqnarray}
&&\mathcal{F}_2(x)\sim \frac{N^6}{\Delta^2}
\int_x^\infty\! d\eta \iint d\epsilon d\epsilon^\prime  \int \frac{dU}{U_0} e^{-U^2/U_0^2}
\nonumber\\
&&\times
\int dq\ e^{i q \eta}
\left\{
1+ \left[\frac{q V_0^2(\epsilon+\epsilon^\prime)}{2\epsilon\epsilon^\prime(\epsilon+\epsilon^\prime+U)}\right]^2\right\}^{-1/2}\nonumber\\
&&\sim \frac{N^6}{\Delta^2}
\int_x^\infty\! d\eta \iint d\epsilon d\epsilon^\prime  \int \frac{dU}{U_0} e^{-U^2/U_0^2}\
 z K_0\left(\eta z\right), \nonumber\\
\label{F2q}
\end{eqnarray}
where
\begin{equation}
z=\frac{2|\epsilon\epsilon^\prime(\epsilon+\epsilon^\prime+U)|}{V_0^2|\epsilon+\epsilon^\prime|}
\label{zdef}
\end{equation}
and $K_0(x)$ is the modified Bessel function of the second kind,
\begin{equation}
z K_0\left(\eta z\right) \simeq \left\{
                                 \begin{array}{ll}
                                   z\ \ln\dfrac{1}{\eta z}, & z\eta \ll 1, \\[0.5cm]
                                   \sqrt{\dfrac{\pi z}{2 \eta}}\ e^{-\eta z}, & z\eta \gg 1.
                                 \end{array}
                               \right.
\label{zK0}
\end{equation}

We first consider the case $U_0\gg V_0$.
The integration over $U$ effectively sets $|U|\sim U_0$ in $z$ and, then, it is sufficient to analyze the contribution of
the domain $V_0<\epsilon$, $V_0<-\epsilon^\prime$, with $U$ replaced by $U_0$.
Since we are interested in resonant processes ($x \gtrsim 1$), we have $\eta \gtrsim 1$. On the other hand,
the contribution of small $|\epsilon|\ll V_0$ and $|\epsilon^\prime|\ll V_0$ to the integral in Eq.~(\ref{F2q})
should be disregarded, since such energy mismatches correspond to $k=1$ resonances.
Therefore, the product $\eta z$ in Eq.~(\ref{zK0}) is typically large and one should then
use the exponential asymptotics
of $K_0(\eta z)$, unless $\epsilon+\epsilon^\prime+U$ is anomalously small.
The main contribution to $\mathcal{F}_2(x)$ turns out to be given by these resonances satisfying $z \eta\ll 1$.
Introducing
$$w=\epsilon-|\epsilon^\prime|,$$
we write
$$z\sim \frac{\epsilon(\epsilon+U_0) w}{V_0^2 U_0}\equiv \frac{y}{\eta}$$
and use the logarithmic asymptotics of Eq.~(\ref{zK0}):
\begin{eqnarray}
\mathcal{F}_2(x)&\sim&
\frac{N^6}{\Delta^2} \int_x \frac{d\eta}{\eta^2}
\int_{V_0} d\epsilon \frac{V_0^2 U_0}{\epsilon(\epsilon+U_0)}
\int_0^{1} dy\ y\
 \ln(1/y) \nonumber
\\
&\sim& \frac{N^6}{x} \frac{V_0^2}{\Delta^2}\ln\frac{U_0}{V_0}.
\label{Frx}
\end{eqnarray}

We see that $\mathcal{F}_2$ is proportional to the logarithmic factor containing a
ratio of the energy shift in the final state due to the diagonal interaction $U_0$
and the off-diagonal matrix element. This logarithmic factor is accumulated in the
energy interval $V_0<\epsilon<U_0$, where the lower bound on $\epsilon$
excludes the appearance of the first-order resonance.
For $U_0\sim V_0$, this logarithmic factor disappears
in $\mathcal{F}_2$, confirming the observation made in Sec.~\ref{s3A4} that there is no
$\ln g$ factor in the resonance condition for $k\sim 1$. Note that, limiting the integration over
$\epsilon$ in Eq.~(\ref{Frx}) by $\mathcal{E}$ from below (with $\mathcal{E} > U_0$) and imposing
the condition of existence of resonances, $\mathcal{F}_2(1) > 1$, one obtains
inequality (\ref{res-2a}) for $\mathcal{E}$.

The structure of the integrals in $\mathcal{F}_2(x)$ suggests, however,
that at higher orders $k\gg 1$, there will be a logarithmic factor in $\mathcal{F}_k(x)$.
Indeed, a typical denominator in the  matrix element (\ref{Vk-exact})
contains a sum of diagonal matrix elements which grows as $\sqrt{k}U_0$
with increasing $k$, while the typical bounds for energies that exclude ``accidental''
resonances in previous generations decrease with $k$. This is expected to
produce a $(\ln k)^{k-1}$ factor after consecutive integration over the ordered set of energies,
leading to the logarithmic shift of the threshold, as discussed in the end of Sec.~\ref{s3A4}.

\section{Many-body delocalization via a resonant subsystem}
\label{App:A}

In this Appendix we consider the model of electronic quantum dot from the perspective that bears a certain similarity with the spin-bath model \cite{prokofiev00}. In the regime $g^{1/3} < N < g^{2/5}$ , the number of direct (first-order) resonances characterizing a typical many-body basis state is given by Eq.~(\ref{p}). Using these resonances, one can proceed up to generation $m$ given by Eq.~(\ref{m}), as explained in Sec.~(\ref{s3A1}). Thus, the original basis state will efficiently mix in this way with
\be
\label{Nr}
{\cal N}_r \sim e^m
\ee
other basis states.

We approximate the whole system as consisting of a resonant subsystem of $m$ well coupled ``pseudospins'' (representing resonances)
 and the remaining electrons.  In the region under consideration the number of resonances is much smaller than the number of electrons. We will argue that, already at $m$ which is only logarithmically larger than unity, the resonant subsystem delocalizes other electrons.

The Hamiltonian is approximated as
\be
\label{Ham}
H = H_r + H^{(0)}_{n} + U_{n} + U_{nr}.
\ee
Here $H_r$ is the full Hamiltonian of the resonant subsystem; it has ${\cal N}_r$
eigenstates which are superpositions of the corresponding ${\cal N}_r$ basis states. In view of the strong delocalization of the resonant subsystem, we will assume that these basis states are strongly mixed (i.e., the inverse participation ratio of a typical exact eigenstate of $H_r$ over the basis states is of the order of $1/{\cal N}_r$.

Further, $H^{(0)}_{n}$ is the noninteracting part of the Hamiltonian of nonresonant electrons; its eigenstates are Slater determinants. Thus, the eigenstates of $H_r + H^{(0)}_{n}$ are assumed to be ``known'': these are the products of nonresonant Slater determinants and the exact states of the resonant subsystem. The remaining two terms are considered as the perturbation: the interaction $U_{n}$ within the nonresonant subsystem and the interaction  $U_{nr}$ between the resonant and nonresonant subsystems.  We want to consider transitions in the nonresonant subsystem assisted by transitions in the resonant system, i.e., a transition from (NI,RI) to (NF,RF), where NI is the initial state of the nonresonant system, RI the initial state of the resonant system, NF the final state of the nonresonant system (different from NI), and RF the final state of the resonant system (different from RI).
The question that we want to answer is under what condition it will be possible to move any given nonresonant electron (let us call it 1) by such processes.

We note first that the level spacing of three-particle states (2,3,4) into which electron 1 could ``decay'' is given by Eq.~(\ref{delta3}), which is larger than typical energies $\sim V \sim \Delta/g$ of the resonant subsystem. Thus, to satisfy the energy conservation, we have to go to the next order and to consider a decay of 1 into 5-particle states. Their level spacing is
$$
\Delta_5(T)\sim\Delta^5/T^4 \sim \Delta^3/E^2,
$$
which is smaller than $\Delta/g$ in the considered energy range. Such a process is possible at second order of perturbation theory, e.g.,  $1 \to (2,3,4)$  and then $4 \to (5,6,7)$.  Here 1,2,3,4,5,6,7 are nonresonant electron/hole states, and both matrix elements originate from $U_{n}$.  This process, however, does not touch yet the resonant subsystem. We thus go to third order and include a matrix element originating from $U_{nr}$. Specifically, let this term be of the form $(7,X) \to (8, X)$, where X is a certain electron state entering one of resonances. We choose this interaction term to be diagonal in X, in order that it has matrix elements between the states of the resonant network. (Moving a resonant electron X to a state Y would normally take the state off the resonant subspace.) It is important that, in spite of the diagonal character ($X \to X$), this interaction has non-diagonal matrix elements between the exact states of the resonant network, since these states are not basis states but rather are formed by resonances.

\begin{figure}
  \centerline{
  \includegraphics[width=5.5cm]{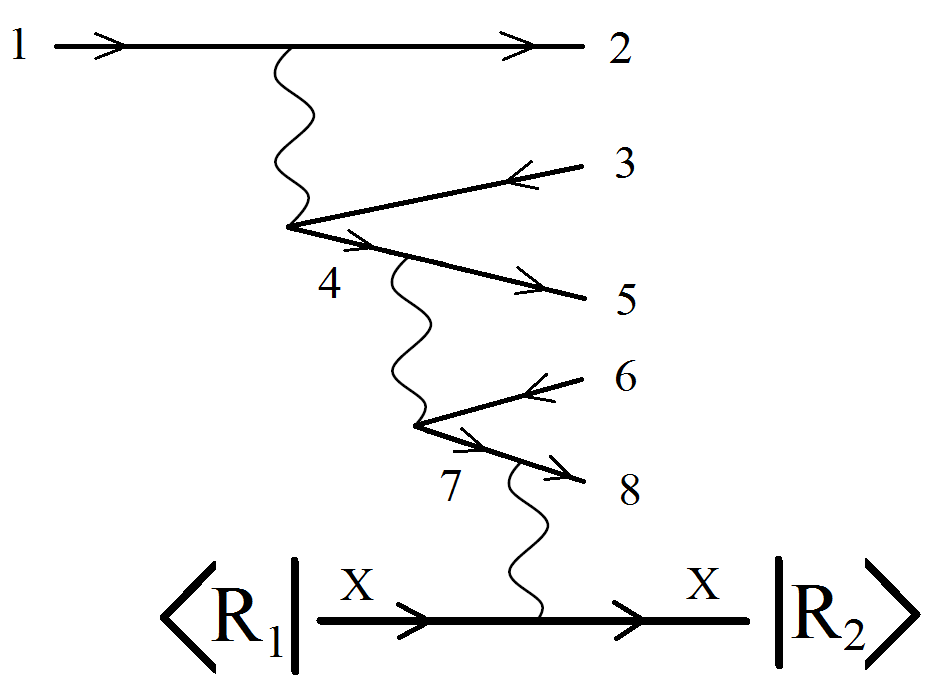}}
  \caption{An example of a third-order decay process stimulated by the resonant subsystem, see text.}
  \label{Fig:App}
\end{figure}

 Drawing the corresponding third-order diagram is straightforward.  Let us now estimate the corresponding matrix element and the relevant level spacing.
 The matrix element of third order is a product of three simple (first-order) matrix elements divided by a product of two energy denominators.
 The first two matrix elements (1,2,3,4) and (4,5,6,7) are of order $V\sim \Delta/g$. The energy denominators are of order $\Delta_3(T)$ each. Thus, the product of these matrix elements divided by the product of energy denominators yields a small dimensionless factor
 \begin{equation}
 \left(\frac{V}{\Delta_3(T)}\right)^2 \sim  \left(\frac{E}{g\Delta}\right)^2.
 \label{A3}
 \end{equation}
 We should check whether it can be compensated by a ratio of the remaining matrix element $M(7,X,8,X)$ to the level spacing. For this purpose, we should estimate the matrix element of $c^\dagger_X c_X$ sandwiched between two different states $R_1$ and $R_2$ of the resonant subsystem. Writing each of these states as a linear combination of ${\cal N}_r$ basis states with random coefficients of order $1/\sqrt{{\cal N}_r}$, we obtain for this matrix element $1/{\cal N}_r$
 times a sum of ${\cal N}_r$ contributions of order unity and random sign, i.e., in total $1/\sqrt{N_r}$. Thus, we get for the matrix element $M(7,X,8,X)$ sandwiched between $R_1$ and $R_2$ (Fig.~\ref{Fig:App}) the estimate
 \begin{equation}
 \langle R_1\left| M(7,X,8,X) \right|R_2\rangle \sim \frac{\Delta} {g \sqrt{{\cal N}_r}}.
 \label{A4}
 \end{equation}
 On the other hand, the level spacing of the final states is controlled by the level spacing of resonant finite states RF given by
 \begin{equation}
 \Delta_\text{RF}\sim \frac{\Delta}{g{\cal N}_r}.
 \label{A5}
 \end{equation}
 Thus, combining Eqs.~(\ref{A3}), (\ref{A4}), and (\ref{A5}), we obtain a parameter determining the possibility of the process that flips ``spin'' 1 in the form
 $$
 \eta_r\sim \left(\frac{E}{g\Delta}\right)^2 \sqrt{{\cal N}_r}.
 $$
 Equating $\eta_r$ to unity and using Eqs. (\ref{p}),  (\ref{m}), and (\ref{Nr}), we find $p \sim (\log g)^{1/2}$ and thus  the delocalization border
 \be
 \label{E23log12}
 E_c \sim \Delta g^{2/3} (\log g)^{1/2},
 \ee
 or, equivalently,
  \be
 \label{Nc-log}
 N_c \sim g^{1/3} (\log g)^{1/4},
 \ee

 It is worth mentioning that, within the above derivation,  the result (\ref{E23log12}) can be traced back to the following: (i) the exponential dependence of ${\cal N}_r$ on $p$ characteristic of a many-body system; (ii) the $1/\sqrt{{\cal N}_r}$ dependence of the matrix element for the interaction between an electron and a state of the resonant subsystem, and (iii) the $1/{\cal N}_r$ dependence of the level spacing of the resonant subsystem.


\begin{thebibliography}{99}
\bibitem{note-diag-BAA} It is worth pointing out that an attempt to take  the diagonal elements into account within the diagrammatic analysis  of quasiparticle decay processes was undertaken in Ref.~\cite{basko06}, see Sec. 7.3 there. The authors of Ref.~\cite{basko06} missed, however, the fact that the diagonal matrix elements give rise to ladders connecting the electronic lines and that there is a factorial growth of the number of possibilities to insert such ladders in a given skeleton diagram. As a result, they came to an erroneous conclusion that the inclusion of the diagonal (Hartree-Fock-type) matrix elements does not essentially change the result for the localization threshold.
\bibitem{remark1} While Ref.~\cite{gornyi04} introduced the conceptual framework for studying Anderson localization in Fock space of an extended system,
it came to the conclusion that, contrary to Ref.~\cite{fleishman80}, there exists a nonzero (double-exponentially small in $1/T$) conductivity at arbitarily
low $T$ due to higher-order many-body scattering processes. The mechanism of this ``creeping conductivity'' in the glassy phase was, however, based on the error in counting the decay processes, as was pointed out to the authors of Ref.~\cite{gornyi04} by I. Aleiner, B. Altshuler, and D. Basko. This erroneous estimate, which included the factorial factor $k!$, was corrected in Ref.~\cite{gornyi05}, where an exponential growth of the number of the skeleton Feynman diagrams with the order of perturbation theory was found (see also Ref.~\cite{ros15}). It is interesting to notice that the inclusion of the diagonal interactions (spectral diffusion) restores the factorial growth of the number of Feynman diagrams for the quantum-dot problem.
\end{thebibliography}
\end{document}